\documentclass[preprint,1p]{elsarticle}
\journal{Journal of Logical and Algebraic Methods in Programming}
\usepackage{amssymb, amsmath, mathtools}
\usepackage{syntax}
\usepackage{float}
\usepackage{listings}
\usepackage{color}
\usepackage[dvipsnames]{xcolor}
\usepackage{tikz}
\usepackage[ligature, inference]{semantic}
\usepackage[linesnumbered,algoruled,boxed,lined]{algorithm2e}
\usepackage{proof}

\newtheorem{proposition}{Proposition}
\newtheorem{definition}{Definition}
\newtheorem{example}{Example}
\newtheorem{remark}{Remark}
\usepackage{bussproofs}
\usepackage{stackengine}

\definecolor{darkblue}{rgb}{0.0,0.0,0.6}
\definecolor{darkgreen}{rgb}{0.0,0.6,0.0}
\definecolor{darkred}{rgb}{0.6,0.0,0.0}
\usepackage{float}

\definecolor{background}{HTML}{EEEEEE}

\lstdefinestyle{pseudoj}{
language=C,                
numbers=left,
numberstyle={\scriptsize},      
frame=Ltbr,
rulesep=.4pt,
mathescape, 
stringstyle=\color{redb},
keywords=[1]{fi, od, do,end,elseif, return, else, int, wire, bool, if, Boolean, do-together, foreach, endfor, endif, while, to, Let},
keywords=[2]{decl-list, init-list,next-list, wiredef-list, var-st, loc-st, append, assignment},
showstringspaces = false,
escapeinside={/*@}{@*/},
basicstyle=\ttfamily\small,
keywordstyle=[2]\color{darkred}\bf,
keywordstyle=[1]\color{darkblue}\bf,
commentstyle=\color{darkgreen},
alsoletter={-},
captionpos=b, %
breaklines=true,
backgroundcolor=\color{background},
}

\lstdefinestyle{pseudoProm}{
language=Promela,                
numberstyle=\scriptsize,      
frame=Ltbr,
rulesep=.4pt,
mathescape, 
stringstyle=\color{redb},
keywords=[1]{fi, od, do,end,elseif, return, else, int, wire, bool, if, Boolean, do-together, foreach, endfor, endif, while, to, Let, recv, recvAck, send, sendAck, synchRecv},
keywords=[2]{decl-list, init-list,next-list, wiredef-list, var-st, loc-st, append, assignment},
showstringspaces = false,
escapeinside={/*@}{@*/},
basicstyle=\tiny ,
keywordstyle=[2]\color{darkred}\bf,
keywordstyle=[1]\color{darkblue}\bf,
commentstyle=\color{darkgreen},
alsoletter={-},
captionpos=b, %
breaklines=true,
backgroundcolor=\color{background},
}

\usepackage{todonotes}
\usepackage{stmaryrd}
\usetikzlibrary{intersections,patterns,decorations.markings,decorations.pathreplacing,arrows,automata,calc,decorations,fit,positioning,backgrounds}
\usetikzlibrary{positioning}

\usetikzlibrary{fadings}
\usetikzlibrary{arrows,shapes,automata,shadows}
\usetikzlibrary{trees}

\tikzstyle{monitor}=[->,draw=orange,fill=orange]
\tikzstyle{recover}=[->,draw=red,fill=red]
\tikzstyle{continue}=[->,draw=green!50!black,fill=green!50!black]

\tikzstyle{component}=[auto,node distance=1.5cm,thick,
  align=center,inner sep=0cm,minimum size=0pt, transform shape,
  ->,>=stealth',shorten >=1pt,
]{}

\tikzstyle{compbig}=[auto,node distance=1.2cm,thick,
  align=center,inner sep=0cm,minimum size=0pt, transform shape,
  ->,>=stealth',shorten >=1pt,
]{}
\tikzstyle{exportsend}=[circle, fill=black!30,minimum size=3.75mm,inner sep=0mm,  label={center:s}]{}
\tikzstyle{exportasend}=[circle, draw, dotted, fill=black!30,minimum size=3.75mm,inner sep=0mm, label={center:as}]{}
\tikzstyle{exportssend}=[circle, draw, fill=black!30,minimum size=3.75mm,inner sep=0mm, label={center:ss}]{}
\tikzstyle{exportreceive}=[rectangle, draw, dotted, fill=black!40,minimum size=3.75mm,inner sep=0mm, label={center:r}]{}

\tikzstyle{bip}=[auto,node distance=1cm,line width=2pt,>=to,thick,align=center]{}

\newcommand{\component}[4]{
  \node[rectangle,draw] (#1) #3 [#2] {
    \\ \\
    \begin{tikzpicture}[component]
      \tikzset{
      place/.style={state,fill=white!89!black,draw=white!50!black,minimum size=0.6cm,inner sep=0cm},
      prime/.style={place,fill=white!95!black,draw=white!70!black},
      spec/.style={state,fill=white!100!black,draw=white!100!black,minimum size=0.6cm,inner sep=0cm}

    }
      #4
    \end{tikzpicture} 
  }

}

\newcommand{\compbig}[4]{
  \node[rectangle,draw] (#1) #3 [#2] {
    \begin{tikzpicture}[compbig]
      \tikzset{
      place/.style={state,fill=white!89!black,draw=white!50!black,minimum size=0.6cm,inner sep=0cm},
      prime/.style={place,fill=white!95!black,draw=white!70!black},
      spec/.style={state,fill=white!100!black,draw=white!100!black,minimum size=0.6cm,inner sep=0cm}

    }
      #4
    \end{tikzpicture} 
  }

}

\pgfmathtruncatemacro\distance{1}

\newcommand{\defref}[1]{Definition~\ref{#1}}
\newcommand{\figref}[1]{Figure~\ref{#1}}

\newcommand{\secref}[1]{Section~\ref{#1}}

\makeatletter
\newcommand{\xdashrightarrow}[2][]{\ext@arrow 0359\rightarrowfill@@{#1}{#2}}
\newcommand{\xdashleftarrow}[2][]{\ext@arrow 3095\leftarrowfill@@{#1}{#2}}
\newcommand{\xdashleftrightarrow}[2][]{\ext@arrow 3359\leftrightarrowfill@@{#1}{#2}}
\def\rightarrowfill@@{\arrowfill@@\relax\relbar\rightarrow}
\def\leftarrowfill@@{\arrowfill@@\leftarrow\relbar\relax}
\def\leftrightarrowfill@@{\arrowfill@@\leftarrow\relbar\rightarrow}
\def\arrowfill@@#1#2#3#4{%
  $\m@th\thickmuskip0mu\medmuskip\thickmuskip\thinmuskip\thickmuskip
   \relax#4#1
   \xleaders\hbox{$#4#2$}\hfill
   #3$%
}
\makeatother

\newcommand{\stend}{\mathtt{end}}
\newcommand{\ststart}{\mathtt{start}}

\newcommand{\synch}[1][\hspace*{5mm}]{\xrightarrow{#1}}

\newcommand{\synchsss}[1][\hspace*{2mm}]{\xrightarrow{#1}}

\lstdefinelanguage{lang} {
	morekeywords = {seq, par, xor, for, end},
	emph={CH}, 
	sensitive=false,
}

\definecolor{pblue}{rgb}{0.13,0.13,1}
\definecolor{pgreen}{rgb}{0,0.5,0}
\definecolor{pred}{rgb}{0.9,0,0}
\definecolor{pgrey}{rgb}{0.46,0.45,0.48}

\makeatletter
\lst@InstallKeywords k{attributes}{attributestyle}\slshape{attributestyle}{}ld
\makeatother

\lstdefinestyle{Grammar}
{language=bash,
keywordstyle=\color{blue},
basicstyle=\ttfamily\normalsize,
morekeywords={end},
morekeywords=[2]{jaber@jaber:},
escapeinside={/*@}{@*/},
mathescape,
numbers=none,
keywordstyle=[2]{\color{red}},
literate={::=}{{\textcolor{red}{::=}}}3
}

\newcommand{\true}{\mathtt{true}}

\newcommand{\buffer}{\mathtt{buff}}
\newcommand{\ctype}{\mathtt{ctype}}
\newcommand{\dtype}{\mathtt{dtype}}
\newcommand{\var}{\mathtt{var}}
\newcommand{\guard}[0]{\mathtt{guard}}
\newcommand{\ufct}[0]{\mathtt{ufct}}

\newcommand{\synchs}{\mathtt{ss}}
\newcommand{\asynchs}{\mathtt{as}}
\newcommand{\recv}{\mathtt{r}}
\newcommand{\internal}{\mathtt{in}}

\newcommand{\semantics[1]}{\lbrc #1 \rbrc}
\newcommand{\init}{\mathtt{init}}
\newcommand{\initit}{\mathit{init}}
\newcommand{\context}{\mathtt{context}}
\newcommand{\st}{\mathtt{ch}}
\newcommand{\stC}{\mathcal{C} }
\newcommand{\while}{\mathtt{while}}
\newcommand{\controlsend}{\mathtt{cs}}
\newcommand{\controlrcv}{\mathtt{cr}}
\newcommand{\unionst}{\mathtt{union}}

\newcommand{\exchannel}[2]{\mathtt{#1 #2}}
\newcommand{\copyport}{\mathtt{copy}}


\newcommand{\xdasharrow}[2][->]{
\tikz[baseline=-\the\dimexpr\fontdimen22\textfont2\relax]{
\node[anchor=south,font=\scriptsize, inner ysep=1.5pt,outer xsep=2.2pt](x){#2};
\draw[shorten <=3.4pt,shorten >=3.4pt,dashed,#1](x.south west)--(x.south east);
}
}

\newcommand{\xarrownormal}[2][->]{
\tikz[baseline=-\the\dimexpr\fontdimen22\textfont2\relax]{
\node[anchor=south,font=\scriptsize, inner ysep=1.5pt,outer xsep=2.2pt](x){#2};
\draw[shorten <=3.4pt,shorten >=3.4pt,#1](x.south west)--(x.south east);
}
}
\newcommand{\synchronous}{\xarrownormal[->,>=latex]{\hspace*{0.4cm}}}

\newcommand{\nat}[0]{\ensuremath{\mathbb{N}}}

\newcommand{\rulename}[1]{{\small \texttt{(#1)}}}

\newcommand{\blue}[1]{\textcolor{NavyBlue}{#1}}
\newcommand{\chtt}[0]{\mathtt{ch}}
\newcommand{\niltt}[0]{\mathtt{nil}}
\newcommand{\sndtt}[0]{\mathtt{snd}}
\newcommand{\sndit}[0]{\mathit{snd}}
\newcommand{\rcvstt}[0]{\mathtt{rcv\_list}}
\newcommand{\rcvsit}[0]{\mathit{rcv\_list}}
\newcommand{\conttt}[0]{\mathtt{cont\_list}}

\newcommand{\prit}[0]{\mathit{pr}}
\newcommand{\ttwhile}[0]{\mathtt{while}}
\newcommand{\ttend}[0]{\mathtt{end}}
\newcommand{\itpr}[0]{\mathit{pr}}
\newcommand{\itpsas}[0]{\mathit{psas}}
\newcommand{\commentgrammar}[1]{\textit{\# #1}}
\newcommand{\chit}[0]{\mathit{ch}}

\newcommand{\sendrm}[0]{\mathrm{send}}
\newcommand{\Choreographies}[0]{\mathit{Chors}}

\newcommand{\chortrans}[1]{\xRightarrow[]{#1}}
\newcommand{\chorconf}[0]{\mathit{ChorConf}}
\newcommand{\chorlabel}[0]{\mathit{ChorLab}}
\newcommand{\chorstate}[0]{\mathit{ChorState}}
\newcommand{\producer}{\texttt{P}}
\newcommand{\consumer}{\texttt{C}}
\begin{document}
\title{From Global Choreographies to\\ Provably Correct and Efficient Distributed Implementations}
\author[aub]{Mohamad Jaber\corref{cor1}}
\ead{mj54@aub.edu.lb}

\author[fr]{Yli\`es Falcone}
\ead{ylies.falcone@univ-grenoble-alpes.fr}

\author[aub]{Paul Attie}
\ead{pa07@aub.edu.lb}

\author[aub]{Al-Abbass Khalil}
\ead{aak103@mail.aub.edu@aub.edu.lb}
\author[aub]{Rayan Hallal}
\ead{rah74@aub.edu.lb}
%
%
\address[aub]{Computer Science Department, American University of Beirut, Beirut, Lebanon}
\address[fr]{Univ. Grenoble Alpes, CNRS, Inria, Grenoble INP, LIG, 38000 Grenoble, France}
%
%
%
\begin{abstract}
We define a method to automatically synthesize provably-correct efficient distributed implementations from high-level global choreographies.
A global choreography describes the execution and communication logic between a set of provided processes which are described by their interfaces.
The operations at the level of choreographies include multiparty communications, choice, loop, and branching.
Choreographies are master-triggered, that is each choreography has one master to trigger its execution. This allows to automatically generate conflict free distributed implementations without controllers.
The behavior of the synthesized implementations follows the behavior of choreographies.
In addition, the absence of controllers ensures the efficiency of the implementation and reduces the communication needed at runtime.
Moreover, we define a translation of the distributed implementations to equivalent \texttt{Promela} versions.
The translation allows verifying the distributed system against behavioral properties.
We implemented a Java prototype to validate the approach and applied it to automatically synthesize micro-services architectures.
We illustrate our method on the automatic synthesis of a verified distributed buying system.
\end{abstract}
\maketitle              
%

\section{Introduction}
%
%
Developing correct distributed software is notoriously difficult.
This is mainly due to their complex structure that consists of interactions between distributed processes.
We mainly distinguish two possible directions to cope with the complexity of the interaction model: (1) high-level modeling frameworks~\cite{BonakdarpourBJQS12}; (2) session types~\cite{BejleriY09,HondaYC08,BonelliC07,VallecilloVR06,GayVRGC10,CharalambidesDA16}.
The former facilitates expressing the communication models but makes efficient code generation difficult.
High-level and expressive communication models require the generation of controllers to implement their communication logic.
For instance, if we consider multiparty interactions with non-deterministic behavior that may introduce conflicts between processes, such conflicts would be resolved by creating new processes (controllers).
Additionally, it is easier to develop distributed systems by reasoning about the global communication model and not local processes.
For these reasons, session types were introduced.
Session types feature the notions of (i) \emph{global protocol} which describes the communication protocol between processes and (ii) \emph{local types} which are the projections of the global protocol on processes.
Session types are generally developed following the below steps:
\begin{enumerate}
\item
design of the global protocol;
\item
automatic synthesis of the local types;
\item
development of the code of processes;
\item
static type checking of the local code of the processes w.r.t. their local protocols.
\end{enumerate}
As a result, the obtained distributed software follows the stipulated global protocol.
However, the current approach to developing session types suffers from several classical problems.
First, there is redundancy in the code of local processes.
Second, the communication logic is tangled as modifying the global protocol requires reimplementing some of the local code of the affected processes. 
Moreover, it suffers from the absence of providing facilities to handle
and combine both communication and computation concerns.
\paragraph{Contributions}
In this paper, we introduce a new framework which allows the automatic synthesis of the local code of the processes starting from a global choreography.
First, inspired from the Behavior Interaction Priority framework (BIP)~\cite{bip2}, we consider a set of components/processes with their interfaces and a configuration file that defines the variables of each component as well as the mapping between ports and their computation blocks.
Then, given a global choreography, which is defined on the set of ports of the components and models coordination and composition operators, we automatically synthesize the local code of the processes that embed all communication and control flow logic.
The choreography allows to define: (1) multiparty interaction; (2) branching; (3) loop; (4) sequential composition; and (5) parallel composition.
Without loss of generality, as in most distributed system applications, we consider master-based protocols.
In master-based protocol, each interaction has a master component deciding whether it can take place and the components involved in the interaction.
This allows for the generation of fully distributed implementations, i.e., without the need of controllers, hence reducing the need for communication at runtime.
Moreover, the synthesized implementations are provably correct, that is we prove that the behavior of the synthesized implementations follows the semantics of choreographies.
Furthermore, we define a translation of the distributed implementations to equivalent \texttt{Promela} versions.
Such a translation allows to verify user-defined properties on the implementations.
Such translation allows to use the SPIN model-checker to verify properties 
Our transformations are implemented in a Java tool that we applied to automatically synthesize micro-service architecture starting from global protocols. 
\paragraph{Differences with HPC 4PAD paper}
This paper revises and extends a paper that appeared in the proceedings of the International Symposium on Formal Approaches to Parallel and Distributed Systems (HPCS 4PAD 2018)~\cite{hpcs4padjaber}.
The additional contributions can be summarized as follows.
First, we define a translation of the distributed implementations to equivalent \texttt{Promela} processes.
This permits the verification of the implementations against (safety and liveness) behavioral properties and thus provides additional confidence in the behavior of the distributed implementation.
Second, we added a synthesis example of a micro-service for a buying system, inspired from the examples tackled in collaboration with Murex Services S.A.L. industry~\cite{murex}.
Third, we revisited and extended the related work.
Finally, we improve the presentation and readability by adding more details and examples.
\paragraph{Paper organization}
The remainder of this paper is structured as follows.
\secref{sec:notation} fixes some notation used throughout the paper.
\secref{sec:preliminary} introduces some preliminary notions, common to choreography and distributed component-based systems.
To illustrate our approach, we present a toy example of a variant of producer-consumer in \secref{sec:illustratingexample}.
In \secref{sec:choreography}, we define the syntax and the semantics of the choreography model.
In \secref{sec:model}, we introduce a distributed component-based model that is used to define the semantics of our choreography model.
In \secref{sec:transformation}, we transform choreographies to distributed component-based systems.
In \secref{sec:correctness}, based on the semantics of choreographies and distributed systems, we show that the transformation in \secref{sec:transformation} is correct in that the distributed system obtained from a choreography produces the same result as the choreography.
In \secref{sec:code}, we provide an efficient code generation of the obtained distributed component-based model and present a real case study. 
In \secref{sec:case}, we present one of the case studies on a micro-service architecture to automatically derive the skeleton of each micro-service, in collaboration with Murex Services S.A.L. industry~\cite{murex}.
%
%
In \secref{sec:promela}, we define a translation of the code generated from a choreography into \texttt{Promela} for the purpose of verifying the generated code. 
We present related work in \secref{sec:rw}.
Finally, we draw conclusions and outline future work in \secref{sec:conclusion}.
%
\section{Notation}
\label{sec:notation}
%
We denote by $\nat$ the set of natural numbers with the usual total orders $\leq$ and $\geq$~; $\nat^*$ denotes the set $\nat \setminus \{0\}$.
Given two natural numbers $a$ and $b$ such that $a \leq b$, we denote by $[a, b]$, the interval between $a$ and $b$, i.e., the set $\{ x \in \nat \mid x \geq a \wedge x \leq b \}$.
A sequence of elements over a set $E$ of length $n \in \nat$ is formally defined as a (total) function from $[1,n]$ to $E$. 
The empty sequence over $E$ (function from $\emptyset$ to $E$) is denoted by $\epsilon_E$ (or $\epsilon$ when clear from the context).
The length of a sequence $s$ is denoted by $|s|$.
The set of (finite) sequences over $E$ is denoted by $E^*$.
The (usual) concatenation of a sequence $s'$ to a sequence $s'$ is the sequence denoted by $s \cdot s'$.
Given two sets $E$ and $F$, we denote by $[E \rightarrow F]$ the set of functions from $E$ to $F$.
Given some function $f \in [E \rightarrow F]$ and an element $e \in E$, we denote by $f(e)$ the element in $F$ associated with $e$ according to $f$.

\section{Preliminary Notions}
\label{sec:preliminary}
%
%
\newcommand{\data}[0]{\ensuremath{\mathit{Data}}}
\newcommand{\datatypes}[0]{\ensuremath{\mathit{DataTypes}}}
\newcommand{\porttypes}[0]{\ensuremath{\mathit{PortTypes}}}
\newcommand{\vars}[0]{\ensuremath{\mathit{Vars}}}
\newcommand{\expr}[0]{\ensuremath{\mathit{Expr}}}
\newcommand{\neutraldata}[0]{\ensuremath{\bot_{\rm d}}}
\newcommand{\neutralexpression}[0]{\ensuremath{\bot_{\rm e}}}
\newcommand{\tuple}[1]{\ensuremath{\left( #1 \right)}}

To later construct a system, we assume an architecture with $n$ components $\{ B_i \}_{i = 1}^{n}$, with $n \in \nat^*$.
At this stage, components are just interfaces with ports for communication.
To each port of a component is attached a (unique) variable.
In this section, we define these notions common to choreographies and component-based systems, later defined in \secref{sec:choreography} and \secref{sec:model} respectively.
\paragraph{Types, variables, expressions, and functions}
We use a set of data types, $\datatypes$, including the set of usual types found in programming languages $\{\mathtt{int}, \mathtt{str}, \mathtt{bool}, \ldots\}$ and a set of (typed) variables $\vars$.
Variables are partitioned over components, i.e., $\vars = \bigcup_{i = 1}^{n} \vars_i$ and $\forall i, j \in [1,n]: i \neq j \implies \vars_i \cap \vars_j = \emptyset$.
Variables take value in a general data domain $\data$ containing all values associated with the types in $\datatypes$ plus a neutral communication element denoted by $\neutraldata$.
We call valuation any function with codomain $\data$.
Moreover, for two valuations $v$ and $v'$, $v'/v$ denotes the valuation where values in $v'$ have priority over those in $v$.
For a set of variables $X \subseteq \vars$, we denote by ${\cal G}(X)$ (resp. $\expr(X)$) the set of boolean (resp. all, i.e., boolean and arithmetic) expressions over $X$, constructed in the usual manner.
Expressions can be used as function descriptions, and, for an expression $e \in \expr(X)$ and a valuation $v \in [ X \rightarrow \data]$, we note $e(v)$ the value in $\data$ of expression $e$ according to $v$. 
\paragraph{Types and ports}
We define the notion of port type, and then of port.
\begin{definition}[Port type]
The set of port types, denoted by $\porttypes$, is $\{\synchs, \asynchs, \recv , \internal \}$, where $\synchs$ (resp. $\asynchs, \recv , \internal$) denotes a synchronous send (resp. asynchronous send, receive, internal) communication type.
\end{definition}
\begin{definition}[Port]
A synchronous send, asynchronous send or internal port is a tuple $(p, x_p, \mathit{dtype}, \mathit{ctype})$ where:
$p$ is the port identifier;
$x_p \in \vars$ is the port variable;
$\mathit{dtype} \in \datatypes$ is the port data type; and
$\mathit{ctype} \in \porttypes$ is the port communication type.
Similarly, a receive port is a tuple $(p, x_p, \mathit{dtype}, \mathit{ctype}, \mathit{buff})$ where $\mathit{buff} \in \data^*$ is the port buffer (used to store values).
\end{definition}
Ports are referred to by their identifier.
In the rest of the paper, we use the dot notation:
\begin{itemize}
	\item
	for a (a)synchronous send or internal port $(p, x_p, \mathit{ptype}, \mathit{ctype})$ or a receive port $(p, x_p, \mathit{ptype}, \mathit{ctype}, \mathit{buff})$, $p.\var$ (resp. $p.\dtype$, $p.\ctype$, $p.\buffer$) refers to $x_p$ (resp. $\mathit{dtype}$, $\mathit{ctype}$, $\mathit{buff}$);
	\item
	for a set of ports $P$, $P.\var$ denotes $\{ p.\var \mid p \in P\}$, the set of variables of the ports in $P$.
\end{itemize}
Given a port $p$, we define the predicate $\mathtt{isSSend}(p)$ (resp., $\mathtt{isASend}$, $\mathtt{isRecv}$, $\mathtt{isInternal}$) that holds true iff (the communication type of) $p$ is a synchronous send (resp., asynchronous send, receive, internal) port, i.e., iff $p.\ctype = \synchs$ (resp. $\asynchs, \recv , \internal$).

To later construct a system, we assume a set of ports ${\cal P}$ and a partition of the ports over components: ${\cal P} = \cup_{i = 1}^{n} P_i$.
We define ${\cal P}^{\rm ss} = \{ p \in {\cal P} \mid \mathtt{isSSend}(p)\}$ (resp. ${\cal P}^{\rm as} = \{ p \in {\cal P} \mid \mathtt{isASend}(p)\}$, 
${\cal P}^{\rm r} = \{ p \in {\cal P} \mid \mathtt{isRecv}(p)\}$) to be the set of all synchronous send port (resp. asynchronous send ports, receive ports) of the system.
Moreover, we denote by ${\cal P}_i^{\rm ss}$ (resp. ${\cal P}_i^{\rm as}$, ${\cal P}_i^{\rm r}$) the set of all synchronous send (resp., asynchronous send, receive) ports of atomic component $B_i$.
\paragraph{Update functions}
Update functions serve to abstract internal computations performed by atomic components.
\begin{definition}[Update function]
An update function $f$ over a set of variables $X \subseteq \vars $ is a sequence of assignments, where each assignment is of the form $x := \mathit{expr}_X$, where $x \in X$ and $\mathit{expr_X} \in \expr(X)$. 
The set of update functions over $X$ is denoted by ${\cal F}(X)$.
\end{definition}
For an update function $f$ and a valuation $v$, executing $f$ on $v$ yields a new valuation $v'$, noted $v' = f(v)$, such that $v'$ is obtained in the usual way by the successive applications of the assignments in $f$ taken in order and where the right-hand side expressions are evaluated with the latest constructed temporary valuation.
\section{Illustrating Example}
\label{sec:illustratingexample}
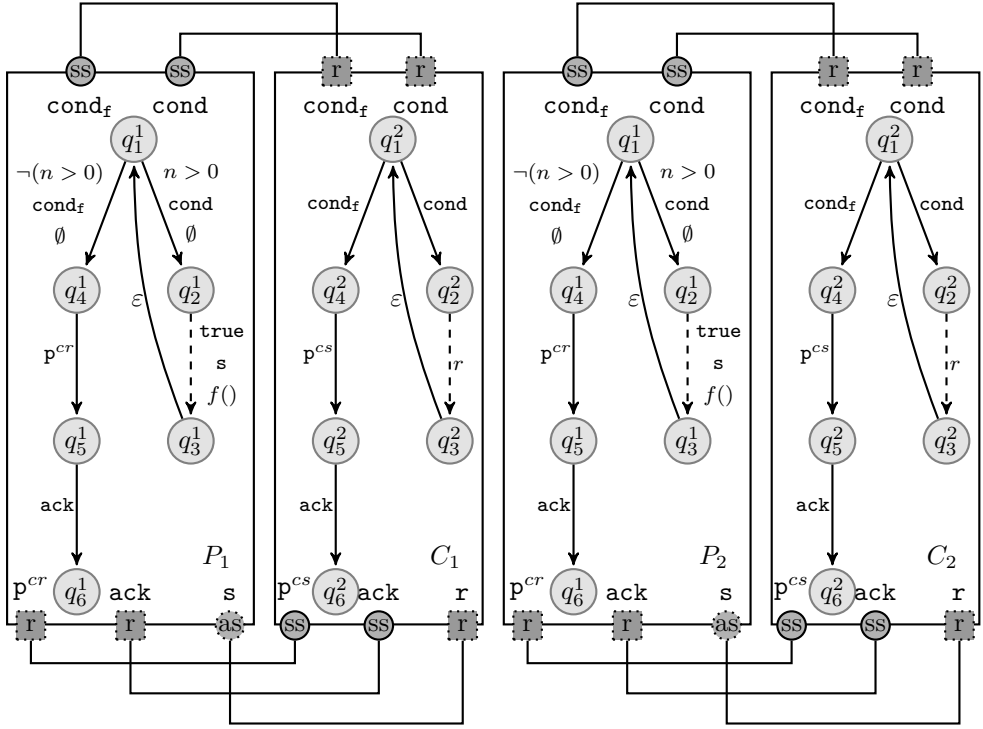
\begin{figure}[t]
    \centering
    \begin{tikzpicture}[bip]
        \component{B1}{minimum size=1cm}{}{
        	\node[place]  (c1l1)  {$q^1_1$};
			\node[place]  (c1l2)  [node distance=2cm,below of = c1l1,xshift=0.75cm]{$q^1_2$};   
			\node[place]  (c1l4)  [node distance=2cm,below of = c1l1,xshift=-0.75cm]{$q^1_4$};   
			\node[place]  (c1l5)  [node distance=2cm,below of = c1l4]{$q^1_5$};   
		\node[place]  (c1l6)  [node distance=2cm,below of = c1l5]{$q^1_6$};   
						\path[->] (c1l4) edge [] node [align=center,left, yshift=0.15cm]   { \footnotesize{$\exchannel{p}{}^{cr}$}} (c1l5);
						
						\path[->] (c1l5) edge [] node [align=center,left, yshift=0.15cm]   {\footnotesize{$\exchannel{ack}{}^{}$}} (c1l6);
						
			\path[->] (c1l1) edge [] node [align=center,right, yshift=0.15cm]   {\footnotesize{$n > 0$}\\ \footnotesize{$\exchannel{cond}{}^{}$} \\ \footnotesize{$\emptyset$}} (c1l2);
			\path[->] (c1l1) edge [] node [align=center, left, yshift=0.15cm]   {\footnotesize{$\lnot (n > 0)$} \\ \footnotesize{$\exchannel{cond_f}{}^{}$} \\ \footnotesize{$\emptyset$}} (c1l4);
			
			\node[place]  (c1l3)  [node distance=2cm,below of = c1l2]{$q^1_3$};   
			
			\path[dashed,->] (c1l2) edge [] node [align=center,right, xshift=0.1cm]   {\footnotesize{$\true$}\\\footnotesize{$\exchannel{s}{}^{}$}\\\footnotesize{$f()$}} (c1l3);
			\draw[bend left=10,->]  (c1l3) to node [auto] {$\varepsilon$} (c1l1);
        };
        \node[exportssend] (p1) at ($(B1.north west)!.3!(B1.north east)$) [label=below:$\exchannel{cond_f}{}^{}$]    {};
        \node[exportssend] (p2) at ($(B1.north west)!.7!(B1.north east)$) [label=below:$\exchannel{cond}{}^{}$]    {};
         \node[exportreceive] (p3) at ($(B1.south west)!.1!(B1.south east)$) [label=above:$\exchannel{p}{}^{cr}$]    {};
        \node[exportreceive] (p4) at ($(B1.south west)!.5!(B1.south east)$) [label=above:$\exchannel{ack}{}^{}$]    {};
                \node[exportasend] (p5) at ($(B1.south west)!.9!(B1.south east)$) [label=above:$\exchannel{s}{}^{}$]    {};
			    \node at ( $(B1.south east)+ (-0.5,0.5)$) [label=above:$P_1$]    {};

       \component{B2}{minimum size=1cm,  right=.25cm of B1}{}{
        	\node[place]  (c2l1)  {$q^2_1$};
			\node[place]  (c2l2)  [node distance=2cm,below of = c2l1,xshift=0.75cm]{$q^2_2$};   
			\node[place]  (c2l4)  [node distance=2cm,below of = c2l1,xshift=-0.75cm]{$q^2_4$};   
			\node[place]  (c2l5)  [node distance=2cm,below of = c2l4]{$q^2_5$};  
			\node[place]  (c2l6)  [node distance=2cm,below of = c2l5]{$q^2_6$};  
			\path[->] (c2l4) edge [] node [align=center,left, yshift=0.15cm]   {\footnotesize{$\exchannel{p}{}^{cs}$}} (c2l5);
			\path[->] (c2l5) edge [] node [align=center,left, yshift=0.15cm]   {\footnotesize{$\exchannel{ack}{}^{}$}} (c2l6);
			
			\path[->] (c2l1) edge [] node [align=center,right, yshift=0.15cm]   {\footnotesize{$\exchannel{cond}{}^{}$}} (c2l2);
			\path[->] (c2l1) edge [] node [align=center, left, yshift=0.15cm]   {\footnotesize{$\quad\exchannel{cond_f}{}^{}$}} (c2l4);
			
			\node[place]  (c2l3)  [node distance=2cm,below of = c2l2]{$q^2_3$};   
			
			\path[dashed,->] (c2l2) edge [] node [label=right:{\footnotesize{$r$}}]   {} (c2l3);
			\draw[bend left=10,->]  (c2l3) to node [auto] {$\varepsilon$} (c2l1);
        };
        \node[exportreceive] (p6) at ($(B2.north west)!.3!(B2.north east)$) [label=below:$\exchannel{cond_f}{}^{}$]    {};
        \node[exportreceive] (p7) at ($(B2.north west)!.7!(B2.north east)$) [label=below:$\exchannel{cond}{}^{}$]    {};
        \node[exportssend] (p8) at ($(B2.south west)!.1!(B2.south east)$) [label=above:$\exchannel{p}{}^{cs}$]    {};
        \node[exportssend] (p9) at ($(B2.south west)!.5!(B2.south east)$) [label=above:$\exchannel{ack}{}^{}$]    {};
                \node[exportreceive] (p10) at ($(B2.south west)!.9!(B2.south east)$) [label=above:$\exchannel{r}{}^{}$]    {};
		    \node at ( $(B2.south east)+ (-0.5,0.5)$) [label=above:$C_1$]    {};	
        
        \draw[-] (p1) -- ++(0,9mm) -| (p6);
        \draw[-] (p2) -- ++(0,5mm) -| (p7);
         \draw[-] (p3) -- ++(0,-5mm) -| (p8);
        \draw[-] (p4) -- ++(0,-9mm) -| (p9);
         \draw[-] (p5) -- ++(0,-13mm) -| (p10);

        \component{B3}{minimum size=1cm, right=.25cm of B2}{}{
        	\node[place]  (c1l1)  {$q^1_1$};
			\node[place]  (c1l2)  [node distance=2cm,below of = c1l1,xshift=0.75cm]{$q^1_2$};   
			\node[place]  (c1l4)  [node distance=2cm,below of = c1l1,xshift=-0.75cm]{$q^1_4$};   
			\node[place]  (c1l5)  [node distance=2cm,below of = c1l4]{$q^1_5$};   
		\node[place]  (c1l6)  [node distance=2cm,below of = c1l5]{$q^1_6$};   
						\path[->] (c1l4) edge [] node [align=center,left, yshift=0.15cm]   { \footnotesize{$\exchannel{p}{}^{cr}$}} (c1l5);
						
						\path[->] (c1l5) edge [] node [align=center,left, yshift=0.15cm]   {\footnotesize{$\exchannel{ack}{}^{}$}} (c1l6);
						
			\path[->] (c1l1) edge [] node [align=center,right, yshift=0.15cm]   {\footnotesize{$n > 0$}\\ \footnotesize{$\exchannel{cond}{}^{}$} \\ \footnotesize{$\emptyset$}} (c1l2);
			\path[->] (c1l1) edge [] node [align=center, left, yshift=0.15cm]   {\footnotesize{$\lnot (n > 0)$} \\ \footnotesize{$\exchannel{cond_f}{}^{}$} \\ \footnotesize{$\emptyset$}} (c1l4);
			
			\node[place]  (c1l3)  [node distance=2cm,below of = c1l2]{$q^1_3$};   
			
			\path[dashed,->] (c1l2) edge [] node [align=center,right, xshift=0.1cm]   {\footnotesize{$\true$}\\\footnotesize{$\exchannel{s}{}^{}$}\\\footnotesize{$f()$}} (c1l3);
			\draw[bend left=10,->]  (c1l3) to node [auto] {$\varepsilon$} (c1l1);
        };
        \node[exportssend] (p1) at ($(B3.north west)!.3!(B3.north east)$) [label=below:$\exchannel{cond_f}{}^{}$]    {};
        \node[exportssend] (p2) at ($(B3.north west)!.7!(B3.north east)$) [label=below:$\exchannel{cond}{}^{}$]    {};
         \node[exportreceive] (p3) at ($(B3.south west)!.1!(B3.south east)$) [label=above:$\exchannel{p}{}^{cr}$]    {};
        \node[exportreceive] (p4) at ($(B3.south west)!.5!(B3.south east)$) [label=above:$\exchannel{ack}{}^{}$]    {};
                \node[exportasend] (p5) at ($(B3.south west)!.9!(B3.south east)$) [label=above:$\exchannel{s}{}^{}$]    {};
		    \node at ( $(B3.south east)+ (-0.5,0.5)$) [label=above:$P_2$]    {};

       \component{B4}{minimum size=1cm,  right=.25cm of B3}{}{
        	\node[place]  (c2l1)  {$q^2_1$};
			\node[place]  (c2l2)  [node distance=2cm,below of = c2l1,xshift=0.75cm]{$q^2_2$};   
			\node[place]  (c2l4)  [node distance=2cm,below of = c2l1,xshift=-0.75cm]{$q^2_4$};   
			\node[place]  (c2l5)  [node distance=2cm,below of = c2l4]{$q^2_5$};  
			\node[place]  (c2l6)  [node distance=2cm,below of = c2l5]{$q^2_6$};  
			\path[->] (c2l4) edge [] node [align=center,left, yshift=0.15cm]   {\footnotesize{$\exchannel{p}{}^{cs}$}} (c2l5);
			\path[->] (c2l5) edge [] node [align=center,left, yshift=0.15cm]   {\footnotesize{$\exchannel{ack}{}^{}$}} (c2l6);
			
			\path[->] (c2l1) edge [] node [align=center,right, yshift=0.15cm]   {\footnotesize{$\exchannel{cond}{}^{}$}} (c2l2);
			\path[->] (c2l1) edge [] node [align=center, left, yshift=0.15cm]   {\footnotesize{$\quad\exchannel{cond_f}{}^{}$}} (c2l4);
			
			\node[place]  (c2l3)  [node distance=2cm,below of = c2l2]{$q^2_3$};   
			
			\path[dashed,->] (c2l2) edge [] node [label=right:{\footnotesize{$r$}}]   {} (c2l3);
			\draw[bend left=10,->]  (c2l3) to node [auto] {$\varepsilon$} (c2l1);
        };
        \node[exportreceive] (p6) at ($(B4.north west)!.3!(B4.north east)$) [label=below:$\exchannel{cond_f}{}^{}$]    {};
        \node[exportreceive] (p7) at ($(B4.north west)!.7!(B4.north east)$) [label=below:$\exchannel{cond}{}^{}$]    {};
        \node[exportssend] (p8) at ($(B4.south west)!.1!(B4.south east)$) [label=above:$\exchannel{p}{}^{cs}$]    {};
        \node[exportssend] (p9) at ($(B4.south west)!.5!(B4.south east)$) [label=above:$\exchannel{ack}{}^{}$]    {};
                \node[exportreceive] (p10) at ($(B4.south west)!.9!(B4.south east)$) [label=above:$\exchannel{r}{}^{}$]    {};
		    \node at ( $(B4.south east)+ (-0.5,0.5)$) [label=above:$C_2$]    {};	
        
        \draw[-] (p1) -- ++(0,9mm) -| (p6);
        \draw[-] (p2) -- ++(0,5mm) -| (p7);
         \draw[-] (p3) -- ++(0,-5mm) -| (p8);
        \draw[-] (p4) -- ++(0,-9mm) -| (p9);
         \draw[-] (p5) -- ++(0,-13mm) -| (p10);

    \end{tikzpicture}
        \caption{A toy example of  a variant of producer-consumer.}
        \label{fig:toy}
\end{figure}

%
%
To illustrate our approach, we consider a toy example of a variant of producer-consumer.
In this section, we illustrate choreographies and their semantics (\secref{sec:choreography}), component-based distributed implementations (\secref{sec:model}) and the synthesis of distributed implementations from choreographies (\secref{sec:transformation}).
\paragraph{Choreography}
The system consists of two components: a producer (\producer{}) and a consumer (\consumer{}).
Initially, \producer{} has a certain number B of messages to send asynchronously through its interface $s$.
%
%
The number of messages that remain to be send is stored in variable $n$ of port $p$.
\producer{} sends its messages asynchronously through interface $s$ and \consumer{} receives messages through interface $r$.
While \producer{} has messages to send ($n > 0$), it applies some computation function $f$ on the message and decrements the value of $n$.
After \producer{} has finished sending, \consumer{} sends an acknowledgment message to \producer{}. We consider two instances of producers (resp. consumers)  $\producer{}_1$ and $\producer{}_2$ (resp. $\consumer{}_1$ and $\consumer{}_2$), where the two pairs are running in parallel.
Below is the choreography modeling the above scenario and realises the transmission of message from \producer{} to \consumer{}.
\begin{align*}
(\mathtt{\while(P_1.cond[n > 0,\emptyset]) \{P_1.s[\true, f()] \synchronous \{ C_1.r[\emptyset] \} \} \bullet C_1.ack \synchronous \{P_1.ack\}}) \\  \parallel 
(\mathtt{\while(P_2.cond[n > 0,\emptyset]) \{P_2.s[\true, f()] \synchronous \{ C_2.r[\emptyset] \} \} \bullet C_2.ack \synchronous \{P_2.ack\}})
\end{align*}
\paragraph{Synthesized distributed system}
The corresponding distributed component-based model is depicted in \figref{fig:toy}.
The system is composed of four components.
Component $\producer{}_1$ has three basic interfaces $\exchannel{ack}{}^{}$ (for receive), $\exchannel{s}{}^{}$ (asynchrnous send) and $\exchannel{cond}{}^{}$ (synchronous cond).
Two other interfaces are generated for control: $\exchannel{cond_f}{}^{}$ and $\exchannel{p}{}^{cr}$.
Condition $\exchannel{cond_f}{}^{}$ is enabled when the condition of the while does not hold.
$\exchannel{p}{}^{cr}$ is used to implement the sequential primitive ($\bullet$).
The two parallel choreographies are independent and corresponds of the parallel execution of $\producer{}_1$ with $\consumer{}_1$ and $\producer{}_2$ with $\consumer{}_2$.  
As can be noticed, there is no need of controllers and one can use a process or thread for each component. 

\paragraph{Promela model}
From the above description of the distributed implementation, we can synthesize a Promela processes (one per componenent).
Interactions will be modeled as channels in Promela.
\section{Global Choreography}
\label{sec:choreography}
%
In this section, we define the global choreography model.
Recall that components are seen as interfaces and a choreography serves the purpose of coordinating the communications and computations of components.
In choreographies, ports are used with guards and update functions.

We start by defining the syntax and then the semantics of choreographies.
\paragraph{Syntax of choreographies}
We introduce the abstract syntax of the global choreography model.
\begin{figure}[tb]
	\centering
	\[
	\begin{small}
	\begin{array}{rclll}
		\chtt		& ::= 	& 		& \blue{\niltt} & \commentgrammar{empty choreography} \\
					& 		& \mid 	& \sndtt \blue{\synchronous} \blue{\tt \{} \rcvstt \blue{\tt \}} \; \blue{\tt :}\; \blue{\tt \langle } t \blue{\tt \rangle} & \commentgrammar{typed send / receive} \\
					& 		& \mid 	& B \, \blue{\oplus} \, \blue{\tt \{} \conttt \blue{\tt \}} & \commentgrammar{conditional master branching}  \\
					& 		& \mid 	& \blue{\ttwhile} \blue{\tt (} \sndtt \blue{\tt )} \, \chtt \, \blue{\ttend} & \commentgrammar{iterative composition} \\
					& 		& \mid 	& \chtt \, \blue{\bullet} \, \chtt & \commentgrammar{sequential composition} \\
					& 		& \mid 	& \chtt \, \blue{\parallel} \, \chtt & \commentgrammar{parallel composition}\\
		\sndtt 	 	& ::=	& 		& \itpsas \blue{\tt [} g, f \blue{\tt ]} & \commentgrammar{synchronous/asynchronous send ports} \\
					& 		& 		& 								&\commentgrammar{\quad with guard \& update function} \\
		\rcvstt 	& ::= 	& 		& \prit \blue{\tt [} f \blue{\tt ]} \mid \prit \blue{\tt [} f \blue{\tt ]} \blue{\tt,}\, \rcvstt & \commentgrammar{list of receive ports with update function} \\
		\conttt 	& ::=	& 		& \sndtt\, \blue{\tt :}\, \chtt \mid \sndtt\, \blue{\tt :}\, \chtt \blue{\tt ,}\, \conttt & \commentgrammar{list of continuations} \\
		t			& \in	&		& \datatypes & \commentgrammar{types} \\
		B			& \in	&		& \{ B_1, \ldots, B_n\} & \commentgrammar{available components} \\
	\itpsas		 	& \in 	& 		& {\cal P}^{\rm ss} \, \cup \, {\cal P}^{\rm as} & \commentgrammar{synchronous/asynchronous} \\
					& 		& 		& 			& \commentgrammar{ \quad send ports identifiers} \\
	\itpr			& \in 	& 		& {\cal P}^{\rm r} & \commentgrammar{receive ports} \\
		g			& \in 	&		& {\cal G}(X) & \commentgrammar{guards} \\ 
		f 			&	\in	&  		& {\cal F}(X) & \commentgrammar{update function} \\
	\end{array}
	\end{small}
	\]
	\caption{Abstract grammar defining the syntax of the choreography model.}
	\label{fig:abstract-grammar-choreographies}
\end{figure}
\begin{definition}[Abstract syntax of the choreography model]
\label{def:choreo:syntax}
	The abstract grammar in \figref{fig:abstract-grammar-choreographies} defines the syntax of the choreography model.
	We denote by $\Choreographies$ the set of choreographies defined by this grammar.
\end{definition}
The definition of choreographies relies on the previously defined concepts such as update functions in ${\cal F}(X)$, guards in ${\cal G}(X)$, the existing types in $\datatypes$, available components in $\{ B_1, \ldots, B_n\}$, and the various types of ports (synchronous and asynchronous send ports in ${\cal P}^{\rm ss}$ and ${\cal P}^{\rm as}$ and receive ports in ${\cal P}^{\rm r}$).
It also relies on the definitions of send port augmented with guard and update function and lists of receive ports and continuations.
A send port augmented with guard and update function is of the form $\itpsas[g, f]$ where $\itpsas$ is a synchronous or asynchronous send port, $g$ a guard, and $f$ an update function.
In a list of receive ports, each element is of the form $\prit[g]$ where $\prit$ is a receive port identifier and $g$ a guard.
In a list of continuations, each element is of the form $\itpsas \blue{{\tt :}} \chit$ where $\itpsas$ is a synchronous or asynchronous send port and $\chit$ is a choreography.  
We extend the dot notation to choreographies and, for a send or receive port augmented with guard and update function, i.e., of the form $\itpsas[g,f]$ or $\itpr[g]$, we note $\itpsas.\guard$ and $\itpr.\guard$ for $g$ and $\itpsas.\ufct$ for $f$. 

Base choreographies include the empty choreography ($\niltt$) and the send/receive communication primitive.
Send/receive communications are of the form $\sndtt \blue{\synchronous} \blue{\tt \{} \rcvstt \blue{\tt \}} \;\blue{\tt :} $ $ \blue{\tt \langle } T \blue{\tt \rangle}$ where $\sndtt$ is a (synchronous or asynchronous) send port, $\rcvstt$ is a list of receive ports and $\blue{\tt :}\; \blue{\tt \langle } T \blue{\tt \rangle}$ is a type annotation with $T \in \datatypes$.

Composite choreographies include the conditional master branching, the iterative, sequential and parallel compositions.
Conditional master branching are of the form $B \, \blue{\oplus} \, \blue{\tt \{} \conttt \blue{\tt \}}$ where $B$ is a component taking the branching decision and $\conttt$ a list of continuations, that is, a list of choreographies guarded by send ports.
The iterative composition of a choreography $\chit$ is of the form $\blue{\ttwhile} \blue{\tt (} \sndtt \blue{\tt )} \, \chit \, \blue{\ttend}$ where $\sndtt$ defines a send port with a guard and an update function.
The component of the send port guides the loop condition.
Given two choreographies $\chit_1$ and $\chit_2$, the sequential (resp. parallel) composition of $\chit_1$ and $\chit_2$ is noted $\chit_1 \, \blue{\bullet} \, \chit_2$ (resp. $\chit_1 \, \blue{\parallel} \, \chit_2$).
\begin{remark}
Guards are not attached to receive ports so as to always permit the reception of data.
Such a choice also allows for generating more efficient code with less communication overhead, and, as communication are master triggered, it avoids deadlock situations.	
\end{remark}

\paragraph{Typing constraints}
Additionally, for a choreography to be well defined, it should respect the following typing constraints:
\begin{itemize}
\item
In a synchronous/asynchronous send port with guard and update function $\itpsas[g,f]$, the variables used in the guard $g$ should belong to the component of port $\itpsas$. 
\item
In a conditional master branching, the send ports in the continuation list should belong to the component.
\end{itemize}
\paragraph{Semantics of choreographies}
\begin{figure}[tb]
	\centering
	\begin{small}
	\[
	\infer[\rulename{nil}]
		{
		(\niltt, \sigma) \chortrans{\tau} \sigma}
		{
		}
	\]
	\[
	\infer[\rulename{synch-sendrcv}]
		{
		(\sndit[g,f] \synchronous {\tt \{} \rcvsit {\tt \}}, \sigma) \chortrans{\{ \sndit, \prit_1, \ldots, \prit_k \}} f \circ f_k \circ \cdots \circ f_1 \circ \sendrm(\sigma, \sndit, \{ \prit_1, \ldots, \prit_k \})
		}
		{
		\sigma \models g
		&
		\rcvsit = \prit_1[f_1], \ldots, \prit_k[f_k]
		}
	\]
	\[
	\infer[\rulename{asynch-sendrcv-1}]
		{
		(\sndit[g,f] \synchronous {\tt \{ } \rcvsit {\tt \}}, \sigma) \chortrans{\{\sndit\}} ( {\tt \{} \rcvsit {\tt \}} , f \circ \sendrm (\sigma, \sndit, \rcvsit))
		}
		{
		\sigma \models g
		}
	\]
	\[
	\infer[\rulename{asynch-sendrcv-2}]
		{
		({\tt \{} \rcvsit {\tt \}} , \sigma) \chortrans{\{\prit\}} ( {\tt \{} \rcvsit {\tt \}} \setminus \{ \prit[f] \}, f(\sigma))
		}
		{
		\prit[f] \in {\tt \{} \rcvsit {\tt \}}
		}
	\]
	\[
	\infer[\rulename{master-branching}]
		{
		( B \oplus \{ \sndit_1[g_1, f_1] : \chit_1, \ldots, \sndit_k[g_k, f_k] : \chit_k \} ) \chortrans{\{ \sndit_j \}} (\chit_j, f_j(\sigma))
		}
		{
		\sigma \models g_j
		}
	\]
	\[
	\infer[\rulename{iterative-tt}]
		{
		\left( \ttwhile (\sndit[g, f]) \, \chit\, \ttend, \sigma \right) \chortrans{\{ \sndit \}} ( \chit \bullet \ttwhile (\sndit[p, f]) \, \chit\, \ttend, f(\sigma) )
		}
		{
		\sigma \models g
		}
	\]
	\[
	\infer[\rulename{iterative-ff}]
		{
		\left( \ttwhile (\sndit[g, f]) \, \chit\, \ttend, \sigma \right) \chortrans{\tau} \sigma
		}
		{
		\sigma \not\models g
		}
	\]
	\[
	\begin{array}{cc}
	\infer[\rulename{sequential-1}]
		{
		(\chit_1 \bullet \chit_2, \sigma) \chortrans{l_1} (\chit'_1 \bullet \chit_2, \sigma')
		}
		{
		(\chit_1, \sigma) \chortrans{l_1} (\chit'_1, \sigma')
		}
	&
	\infer[\rulename{sequential-2}]
		{
		(\chit_1 \bullet \chit_2, \sigma) \chortrans{l_1} (\chit_2, \sigma')
		}
		{
		(\chit_1, \sigma) \chortrans{l_1} \sigma'
		}	
	\end{array}
	\]
	\[
	\begin{array}{cc}
	\infer[\rulename{parallel-1}]
		{
		(\chit_1 \parallel \chit_2, \sigma_1) \chortrans{l_1} (\chit'_1 \parallel \chit_2, \sigma_1')
		}
		{
		(\chit_1,\sigma_1) \chortrans{l_1} (\chit'_1,\sigma_1')
		}
	&
	\infer[\rulename{parallel-2}]
		{
		(\chit_1 \parallel \chit_2, \sigma_2) \chortrans{l_2} (\chit_1 \parallel \chit'_2, \sigma'_2)
		}
		{
		(\chit_2, \sigma_2) \chortrans{l_2} (\chit'_2, \sigma'_2)
		}	
	\end{array}
	\]
	\[
	\begin{array}{cc}
	\infer[\rulename{parallel-3}]
		{
		(\chit_1 \parallel \chit_2, \sigma_1) \chortrans{l_1} (\chit_2, \sigma'_1)
		}
		{
		(\chit_1, \sigma_1) \chortrans{l_1} \sigma'_1
		}
	&
	\infer[\rulename{parallel-4}]
		{
		(\chit_1 \parallel \chit_2, \sigma_2) \chortrans{l_2} (\chit_1, \sigma'_2)
		}
		{
		(\chit_2, \sigma_2) \chortrans{l_2} \sigma'_2
		}	
	\end{array}
	\]	
	\end{small}
	\caption{Rules defining the transitions in the semantics of choreographies.}
	\label{fig:semantics-choreographies}
\end{figure}
In the following, we consider well-typed choreographies built with the syntax in \defref{def:choreo:syntax}.
We define the (structural operational) semantics of choreographies.
For this, we consider that states of a choreography are valuations of the component variables in $[ X \rightarrow \data]$.
Recall that variables and ports are partitioned over components. 
We denote by $\chorstate$ the set of choreography states.

Before actually defining the semantics, we need to model the effect of communication on the choreography state.
We model the sending through a port to a set of ports with a function $\sendrm : \chorstate \times \left( {\cal P}^{\rm as} \cup {\cal P}^{\rm s} \right) \times 2^{{\cal P}^{\rm r}} \rightarrow \chorstate$ that takes as input a choreography state and outputs a choreography state when a communication occurs from the (synchronous or asynchronous) send port of a component to the receive ports of some components: $\sendrm(\sigma, \sndit, \{ \rcvsit \})$ is state $\sigma$ where the value of variable of port $\sndit$ is used to update the variables attached to ports in $\{ \rcvsit \}$.
Formally: $\sendrm(\sigma, \sndit, \{ \rcvsit \}) = \sigma\left[ \{ \rcvsit \}.\var \mapsto \sigma(\sndit.\var) \right]$, it is state $\sigma$ where we apply the substitution that assigns all the variables in $\{ \rcvsit \}.\var$ to $\sigma(\sndit.\var)$.

We are now able to define the semantics of choreographies.
\begin{definition}[Semantics of choreography model]
\label{def:chor:sem}
The semantics of choreographies is an LTS $\left( \chorconf, \chorlabel, \Rightarrow \right)$ where~:
\begin{itemize}
	\item $\chorconf \subseteq \left( \Choreographies \times \chorstate \right) \cup \chorstate $ is the set of configurations and $\chorstate \subseteq \chorconf$ is the set of final configurations;
	\item $\chorlabel \subseteq \left( 2^{\cal P} \setminus \{\emptyset\} \cup \{ \tau\} \right)$ is the set of label where each label is either a set of ports or label $\tau$ for silent transitions;
	\item $\chortrans{} \subseteq \chorconf \times \chorlabel \times \chorconf$ is the least set of (labelled) transitions satisfying the rules in \figref{fig:semantics-choreographies};
\end{itemize}
\end{definition}
Whenever for two configurations $c, c' \in \chorconf$ and a label $l \in \chorlabel$, $(c, l, c') \in\, \chortrans{}$, we note it $c \chortrans{l} c'$. 
The rules in \figref{fig:semantics-choreographies} can be intuitively understood as follows:
\begin{itemize}
	\item Rule \rulename{nil} states that choreography $\niltt$ terminates in any state $\sigma$ and produces the terminal configuration $\sigma$.
	\item Rule \rulename{synch-sendrcv} describes the synchronous send/receive primitive.
	The component of port $\sndit$ transfers data to the components with the receive ports in $\rcvsit$ whenever the guard $g$ attached to $\sndit$ holds true from the starting state $\sigma$.
	If the list of receive ports (with update functions) is $\prit_1[f_1], \ldots, \prit_k[f_k]$, the choreography terminates in a state obtained after the data transfer defined by $\sendrm(\sigma, \sndit, \{ \prit_1, \ldots, \prit_k \})$ and the applications of the update functions $f, f_1, \ldots, f_k$ of the send and receive ports.
	Note that the application order does not influence the resulting state as these update functions apply to disjoint variables.
	\item Rule \rulename{asynch-sendrcv-1} describes the first part of an asynchronous send/receive primitive.
	As in the synchronous send/receive primitive, the component of port $\sndit$ transfers data to the components with the receive ports in $\rcvsit$ whenever the guard $g$ attached to $\sndit$ holds true from the starting state $\sigma$.
	However, the state of the receiving component is only updated with the transferred data (with $\sendrm(\sigma, \sndit, \{ \prit_1, \ldots, \prit_k \})$) and the receiving components do not apply their update functions.
	\item Rule \rulename{asynch-sendrcv-2} describes the second part of an asynchronous send/receive primitive. 
	A receive port $\prit[f]$ in the list of receive ports to be executed $\rcvsit$ applies the attached updated function $f$ to the current state and is removed from the list of received ports to be executed.
	\item Rule \rulename{master-branching} describes the (conditional) master branching from component $B$ on one of its continuation $\sndit_j[g_j, f_j] : \chit_j$ whenever the guard $g_j$ attached to port $\sndit_j$ holds true.
	The resulting configuration consists of the choreography $\chit_j$ and the state $f_j(\sigma)$ (resulting from the application of the attached update function $f_j$ to $\sigma$).  
	\item Rule \rulename{iterative-tt} describes the first case of the iterative composition of a choreography $\chit$ under the condition $\sndit[g, f]$ (which consists of a send port $\sndit$, a guard $g$, and an update function $f$).
	When $g$ holds true in $\sigma$, the resulting configuration consists of the choreography $\chit$ sequentially composed with the same starting choreography to be executed in state $\sigma$ updated by $f$.
	\item Rule \rulename{iterative-ff} describes the second case of the iterative composition of a choreography $\chit$ under the condition $\sndit[g, f]$.
	When $g$ holds false in $\sigma$, the choreography terminates in the (unmodified) state $\sigma$.
	\item Rules \rulename{sequential-1} and \rulename{sequential-2} describe the possible evolutions of two sequentially composed choreographies $\chit_1$ and $\chit_2$.
	Rule \rulename{sequential-1} describes the case where the execution of choreography $\chit_1$ does not terminate and evolves to a configuration $(\chit_1, \sigma_1')$ which leads to the global configuration $(\chit'_1 \bullet \chit_2, \sigma_1')$.
	Rule \rulename{sequential-2} describes the case where the execution of choreography $\chit_1$ terminates and evolves to a final configuration $\sigma_1'$ which leads to the global configuration $(\chit_2, \sigma_1')$ (where the second choreography $\chit_2$ is to be executed in state $\sigma_1'$).
	\item Rules \rulename{parallel-1} to \rulename{parallel-4} describe the possible evolutions of two choreographies $\chit_1$ and $\chit_2$ composed in parallel.
	Rules \rulename{parallel-1} and \rulename{parallel-2} describe the evolutions where $\chit_1$ performs a computation step and terminates or not.
	Rules \rulename{parallel-3} and \rulename{parallel-4} describe the evolutions where $\chit_2$ performs a computation step.
\end{itemize}
\section{Distributed Component-based Framework}
\label{sec:model}
%
%
In this section, we introduce a component-based framework, inspired from the Behavior Interaction Priority framework (BIP)~\cite{bip2}.
In the BIP framework, atomic components communicate through an interaction model defined on the interface ports of the atomic components.
Moreover, all ports have the same type. 
Unlike BIP, we distinguish between four types of ports: (1) synchronous send; (2) asynchronous send; (3) asynchronous receive; and (4) internal ports.
The new port types allow to (1) easily model distributed system communication models; (2) provide efficient code generation, under some constraints, that does not require to build controllers to handle conflicts between multiparty interactions.
%
\subsection{Atomic Components}
%
Atomic components are the main computation blocks.
Atomic components are endowed with a set of variables used in their computation.
An atomic component is defined as follows. 
\begin{definition}[Atomic component - syntax]
An atomic component $B$ is a tuple $(P,$ $X, L, T)$, where $P$ is a set of ports; $X$ is a set of variables such that $X \subseteq \vars$ and $P.\var \subseteq X$; $L$ is a set of control locations; and $T \subseteq \left( L \times P \times {\cal G}(X) \times {\cal F}(X) \times L \right)$ is a set of transitions.
\end{definition}
Transitions make the system move from one control location to another by executing a port.
Transitions are guarded and are associated with the execution of an update function.
In a transition $(\ell, p, g, f, \ell') \in T$, $\ell$ and $\ell'$ are respectively the source and destination location, $p$ is the executed port, $g$ is the guard, and $f$ is the update function.

The semantics of an atomic component is defined as an LTS.
A state of the LTS consists of a location $\ell$ and valuation $v$ of the variables where a valuation is a function from the variables of the component to a set of values.
The atomic component can transition from state $ \tuple{\ell,v}$ to state $\tuple{\ell',v'}$ using a transition $\tuple{ \ell, p, d, g, f, \ell'} \in T$ if (i) the guard of the transition holds ($g(v)$ holds true) (ii) the application of update function $f$ to valuation $v_{pd}/v$ yields $v'$ where $v_{pd}$ is the valuation associating $p.\var$ with $d \in \data$, which is a value possibly received from other components.
\begin{definition}[Atomic component - semantics]
The semantics of an atomic component $(P, X, L, T)$ is a labelled transition system, i.e., a tuple $( Q, {\cal P} \times \data, \rightarrow)$, where:
\begin{itemize}
	\item
	$Q \subseteq L \times [ X \rightarrow \data ]$ is the set of states,
	\item
	${\cal P} \times \data$ is the set of labels where a label is a pair made of a port and a value, and
	\item
	$\rightarrow \subseteq Q \times P \times \data \times Q$ is the set of transitions defined as:
	\[
	\{\tuple{ \tuple{\ell,v}, \tuple{p, d}, \tuple{\ell',v'}} \mid \exists \tuple{ \ell, p, g, f, \ell'} \in T: g(v) \wedge v' = f(v_{pd}/v) \}.
	\]
\end{itemize}
\end{definition}
When $\left(q, (p, d), q' \right) \in T$, we note it $q \synch[p/d] q'$.
Moreover, we use states as functions: for $x \in X$ and $q = (l, v)$, $q(x)$ is a short for $v(x)$. 

To later construct a system, we shall use a set of $n$ atomic components $\{B_i = (P_i, Q_i, T_i)\}_{i = 1}^{n}$

Synchronization between the atomic components is defined using the notion of interaction. 
\begin{definition}[Interaction]
An interaction from component $B_i$ to components $\{B_j\}_{j \in J}$, where $i \notin J$, is a pair $(p_i, \{p_j\}_{j \in J})$, where:
\begin{itemize}
	\item
	 $p_i$ is its send port (synchronous or asynchronous) that belongs to the send ports of atomic component $B_i$, i.e., $p_i \in {\cal P}_i^{\rm ss} \cup {\cal P}_i^{\rm as}$;
	\item
	$\{p_j\}_{j \in J}$ is the set of receive ports, each of which belongs to the receive ports of atomic component $B_j$, i.e., $\forall j \in J: p_j \in {\cal P}_j^{r}$.
\end{itemize}
\end{definition}
An interaction $(p_i, \{p_j\}_{j \in J})$ is said to be synchronous (resp. asynchronous) iff $\mathtt{isSSend}(p_i)$ (resp. $\mathtt{isASend}(p_i)$) holds.
%
\subsection{Composite Components}
%
A composite component consists of several atomic components and a set of interactions.
The semantics of a composite component is defined as a labeled transition system where the transitions depend on the interaction types.
\begin{figure}[tb]
	\centering
	\begin{small}
	\[
	\infer[\rulename{synch-send}]
		{
		(q_1, \ldots, q_n) \synch[a] (q'_1, \ldots, q'_n)
		}
		{
		\begin{array}{l}
		\mathtt{isSSend}(p_i) \\
		a = (p_i, \{ p_j \}_{j \in J}) \in \gamma \\
		d = q_i(p_i.\var) \in \data
		\end{array}
		&
		\begin{array}{l}
		\forall k \in J \cup \{ i \}: q_k \synch[p_k/d] q'_k\\
		\forall k \notin J \cup \{ i \}: q_k = q'_k\\
		\end{array}
		&
		\begin{array}{l}
		\forall j \in J: q_j(p_j.\buffer) = \epsilon
		\end{array}
		}
	\]
	\[
	\infer[\rulename{asynch-send}]
		{
		(q_1, \ldots, q_n) \synch[a] (q'_1, \ldots, q'_n)
		}
		{
		\begin{array}{l}
		\mathtt{isASend}(p_i) \\
		a = (p_i, \{ p_j \}_{j \in J}) \in \gamma\\
		d = q_i(p_i.\var) \in \data
		\end{array}
		&
		\begin{array}{l}
		\forall k \in J \setminus \{i\}: q'_k = q_k\\
		q_i \synch[p_i/d] q'_i
		\end{array}
		&
		\begin{array}{l}
		\forall j \in J:\\
		q'_j(p_j.\buffer) = q_j(p_j.\buffer) \cdot d
		\end{array}
		}
	\]
	\[
	\infer[\rulename{recv}]
		{
		(q_1, \ldots, q_n) \synch[\tau] (q'_1, \ldots, q'_n)
		}
		{
		\mathtt{isRecv}(p_j)
		&
		\begin{array}{l}
		q_j \synch[p_j/d] q'_j\\
		\forall k \neq j: q_k = q'_k
		\end{array}
		&
		\begin{array}{ll}
		q_j(p_j.\buffer) = d \cdot D & d \in \data \\
		q'_j(p_j.\buffer) = D & D \in \data^*
		\end{array}
		}
	\]	
	\[
	\infer[\rulename{internal}]
		{
		(q_1, \ldots, q_n) \synch[\tau] (q'_1, \ldots, q'_n)
		}
		{
		\mathtt{isInternal}(p_i)
		&
		q_i \synch[p_i/\neutraldata] q'_i
		&
		\begin{array}{l}
		\forall k \neq i: q_k = q'_k
		\end{array}
		}
	\]	
	\end{small}
	\caption{Semantic rules defining the behavior of composite components.}
	\label{fig:composite-rules}
\end{figure}
\begin{definition}[Composite component]
A composite component built over atomic components $B_1, \ldots, B_n$ and parameterized by a set of interactions $\gamma$, noted $\gamma(B_1, \ldots, B_n)$, is defined as a transition system $(Q, \gamma \cup \{ \tau \}, \rightarrow)$, where~:
\begin{itemize}
	\item $Q = \bigotimes_{i=1}^{n} Q_i$ is the set of configurations,
	\item $\gamma \cup \{\tau\}$ is the set of labels which consist of interactions and $\tau$ for silent transitions, and 
	\item $\rightarrow$ is the least set of transitions satisfying the rules in \figref{fig:composite-rules}.
\end{itemize} 
\end{definition}
The semantic rules in \figref{fig:composite-rules} can be intuitively understood as follows:
\begin{itemize}
	\item 
	Rule \rulename{synch-send} describes synchronous interactions, i.e., the interactions of the form $(p_i, \{p_j\}_{j \in J})$ where $\mathtt{isSSend}(p_i)$, where some component $B_i$ synchronously sends to some components $B_j, j \in J$.
	The variable attached to port $p_i$ of $B_i$ ($p_i.\var$) gets evaluated to some value $d \in \data$, which is transmitted.
	All components $B_k$, $k \in J \cup \{ i \}$, perform a transition $q_k \synch[p_k/d] q'_k$, and other components do not move ($q_k = q'_k$ for $k \notin J \cup \{ i \}$).
	The rule requires that all the corresponding receive ports have no pending messages (their buffers are empty, i.e., $\forall j \in J: q_j(p_j.\buffer) = \epsilon$).
	The states of all the involved components are simultaneously updated through the transition $q_k \synch[p_k/d] q'_k$, for $ j \in J \cup \{ i \}$.
	%
	\item
	Rule \rulename{asynch-send} describes asynchronous interactions, i.e., the interactions of the form $(p_i, \{p_j\}_{j \in J})$ where $\mathtt{isSSend}(p_i)$, where some component $B_i$ asynchronously sends to some components $B_j, j \in J$.
	The rule resembles the previous one, except that it does not require the participation of the receiving components.
	Only the sending component performs a transition $q_i \synch[p_i/d] q'_i$ and the receiving components (as well as the other components) do not move.
	Value $d \in \data$ is appended to the buffer of the corresponding receive ports ($\forall j \in J: q'_j(p_j.\buffer) = q_j(p_j.\buffer) \cdot d$).
	\item
	Rule \rulename{recv} describes the autonomous execution of receive port $p_j$ of some component $B_j$.
	The rule requires that the buffer of port $p_j$ is non-empty ($q_j(p_j.\buffer) = d \cdot D$, with $d \in \data$ and $D \in \data^*$).
	The execution of this interaction makes component $B_j$ perform a transition $q_j \synch[p_j/d] q'_j$ and consumes value $d$ in  buffer $p_i.\buffer$.
	\item
	Rule~\rulename{internal} describes the autonomous execution of an internal port $p_i$ of component $B_i$ where only the local state of $B_i$ is updated by performing the transition $q_i \synch[p_i/\bot_d] q'_i$. 
\end{itemize}
Finally, a system is defined as a composite component where we specify the initial states of its atomic components. 
\begin{definition}[System]
A system is a pair $(\gamma(B_1, \ldots, B_n), \init)$, made of a composite component and $\init \in \bigotimes_{i=1}^{n} Q_i$ its initial state. 
\end{definition}
%
%
%
\section{Transformations}
\label{sec:transformation}
%
We start with a composite component consisting of $n$ atomic components $\{ B_1, \ldots, B_n\}$ with their interface ports and variables.
That is, the behaviors of the input atomic components are empty.
Atomic components can be considered as services with their interfaces but with undefined behaviors.

In this section, we define how to automatically synthesize the behaviors of atomic components given a global choreography model.
To realize choreographies as atomic components we follow the syntactic structure of the choreography.
This facilitates the definition of the transformation from choreographies to components and lead to a clearer implementation.
%
\subsection{Preliminary Notions and Notation}
%
We introduce some preliminary concepts and notations that will serve the realization of choreographies as components.
As we are inductively transforming choreographies to components, we need to synchronize the execution of the independently generated choreographies.
For this, we define three auxiliary functions that takes a choreography as input and give the components that:
\begin{itemize}
	\item are involved in the realization of the choreography -- function $\stC$. 
	\item need to be notified for the choreography to start -- function $\ststart$,
	\item need to terminate for the choreography to terminate -- function $\stend$,
\end{itemize}
The definitions of the two latter functions follow from the semantics of choreographies (\defref{def:chor:sem}).
Note, in the following definitions, when referring to a port $p$ with a guard and/or update function involved in a choreography, we note $p[-]$ when the guard and/or update function is irrelevant to the definition.
\paragraph{Function $\stC$}
We define $\stC(\st)$ as the set of indexes of all components involved in choreography $\st$.
\begin{definition}[Function $\stC$]
Function $\stC : Choreographies \rightarrow 2^{[1,n]} \setminus \{ \emptyset \}$ is inductively defined over choreographies as follows:
\[
\begin{array}{rl}
	\stC(\itpsas) = & \{ i \} \text{ if } \exists i \in [1,n] : \itpsas \in {\cal P}^{\rm ss}_i \cup {\cal P}^{\rm as}_i \\
	\stC(\prit[-]) = & \{ i \} \text{ if } \exists i \in [1,n] : \prit \in {\cal P}^{\rm r}_i \\
	\stC(\prit[-], \rcvstt) = & \stC(\prit[-]) \cup \stC(\rcvstt) \\
	\stC(\niltt) = & \emptyset \\
	\stC(\sndtt \synchronous  \{ \rcvstt \}) = & \stC(\sndtt) \cup \stC(\rcvstt) \\       
	\stC(B_i \oplus \{ \conttt \}) = & \{ i \} \cup \stC(\conttt) \\
	\stC\big(\ttwhile(\sndtt) \, \st \,  \ttend \big) = & \stC(\sndtt) \cup \stC(\st) \\
	\stC(\st_1 \bullet \st_2) = & \stC(\st_1) \cup \stC(\st_2) \\
	\stC(\st_1 \parallel \st_2) = & \stC(\st_1) \cup \stC(\st_2)\\
\end{array}
\]
\end{definition}
\paragraph{Function $\ststart$}
We define $\ststart(\st)$ as the set of indexes of the components in $\st$ that should be notified to trigger the start of $\st$.
\begin{definition}[Function $\ststart$]
Function $\ststart : Choreographies \rightarrow 2^{[1, n]} \setminus \{ \emptyset \}$ is inductively defined over choreographies as follows:
\[
\begin{array}{rl}
	\ststart(\niltt) = & \emptyset \\
	\ststart(\sndtt \synchronous  \{ \rcvstt \}) = & \stC(\sndtt) \\       
	\ststart(B \oplus \{ \conttt \} ) = & \stC(B) \\
	\ststart\big(\ttwhile (\sndtt) \, \chtt \,  \ttend \big) = & \stC(\sndtt) \\
	\ststart(\st_1 \bullet \st_2) = & \ststart(\st_1) \\
	\ststart(\st_1 \parallel \st_2) = & \ststart(\st_1) \cup \ststart(\st_2) 
\end{array}
\]	
\end{definition}
Intuitively, to start a simple synchronous or asynchronous send/receive, the component of its corresponding send port should be notified.
Conditional master branching choreographies can be started by notifying their corresponding master component. 
Iterative choreographies can be started by notifying the component of its corresponding send port.
A choreography consisting of the sequential composition of two choreographies can be started by notifying the components that can start the first choreography.
A choreography consisting of the parallel composition of two choreographies can be started by notifying the components that can start the two choreographies of the composition.
\paragraph{Function $\stend$}
Similarly, we define $\stend(\st)$ as the set of indexes of the components involved in $\st$ that need to terminate so that $\st$ terminates.
\begin{definition}[Function $\stend$]
Function $\stend : Choreographies \rightarrow 2^{[1, n]} \setminus \{ \emptyset \}$ is inductively defined over choreographies as follows:
\[
\begin{array}{rl}
	\stend(\niltt) = & \emptyset \\
	\stend(\sndit[-] \synchronous  \{ \rcvstt \}) = & \stC(\rcvstt) \mbox{ if } \sndit \in {\cal P}^{\rm ss} \\       
	\stend(\sndit[-] \synchronous  \{ \rcvstt \}) = & \stC(\sndit) \mbox{ if } \sndit \in {\cal P}^{\rm as} \\       
	\stend(B \oplus \{ \conttt \} ) = & \stC(\conttt) \\
	\stend\big(\ttwhile(\sndtt) \, \st \,  \ttend \big) = & \stC(\sndtt) \\
	\stend(\st_1 \bullet \st_2) = & \stend(\st_2) \\
	\stend(\st_1 \parallel \st_2) = & \stend(\st_1) \cup \stend(\st_2)
\end{array}
\]	
\end{definition}
We consider that a synchronous send/receive is terminated when all the components involved in the sending and receiving ports are terminated.
However, if the send part is asynchronous, any subsequent choreography can start after the sending is complete.
Conditional master branching choreographies are terminated when the corresponding master component has terminated.
Iterative choreographies are terminated when the component of the send port (with its guard used as condition) has terminated.
A choreography consisting of the sequential composition of two choreographies has terminated when the second choreography in the composition has terminated.
A choreography that consists of the parallel composition of two choreographies has terminated when the first and second choreographies have terminated.
\paragraph{Representing components}
In the sequel, we represent receive ports (resp. synchronous send, asynchronous send) using dashed square labeled with $r$ (resp. circle with solid border labeled with $ss$, circle with dashed border labeled with $as$).
We also omit the border for send ports when synchrony is out of context and label it with $s$. 
%
\subsection{Generation of Distributed CBSs}
\label{sec:tdcbs}
%

We consider a global choreography $\st$ defined over the set of ports ${\cal P} = \cup_{i = 1}^{n}P_i$ of a given set of atomic components (with empty behavior) with their corresponding variables.
Given a choreography $\st$, we define a set of transformations that allows to generate the behaviors and the corresponding interactions of the distributed components $S = (B, \init)$.
Moreover, as we progressively build system $S$, we consider that it has a context to denote the current state where a choreography should be appended. 
For this, ${\cal S} = (S, \context)$ denotes a system with its corresponding context where $\context$ is a function that takes an atomic component as input and returns a location, i.e., $\context(B_i) \in L_i$ to denote the current context of atomic components $B_i$.
The building of the final system is done by induction, following the syntactic structure of the input choreography and uses the continuously updated context.
Any step for constructing the component ensures that the context of each component consists of a unique state.

Initially, we consider a system skeleton ${\cal S} = (S, \context)$, where $B = \gamma(B_1, \ldots, B_n)$ with: (1) $\gamma = \emptyset$; (2) $B_i = (P_i, \emptyset, \{ l_i \}, \emptyset)$; (3) $\initit = (l^{\init}_1, \ldots, l^{\init}_n)$; and (4) $\context(B_i) = l^{\init}_i$; for $i \in [1,n]$. 
The initial location of the obtained system remains unchanged, i.e., it is $\initit$.
As such, for the sake of clarity, we omit it in our construction.
Moreover, all variables are initialized to their default value. 
%
\subsubsection{Send/Receive}
%
Send/receive choreography updates the participating components by adding a transition from the current context and labeling it by the corresponding send or receive port from the choreography.
In order to avoid inconsistencies between same ports but from different choreographies, we create a copy of each port of the choreography ($\copyport$).  $\copyport(p)$ is a new port that has the same function and guard, but a different name.
We also add the corresponding interaction between the send and the receive ports.
Finally, we update the context of the participants to be the corresponding new added states.
As such, if the initial context of each component consists of one state, then the resulting system (after applying the send/receive choreography) also guarantees that each of its components also consists of one state. 
Note that an interaction connected to a synchronous send port and receive ports can be considered as a multiparty interaction with a master trigger, which is the send port.
As such, this allows to efficiently implement multiparty interactions. 

\begin{remark}
Creating a copy for each port per choreography is necessary to generate efficient and correct distributed implementation. As for efficiency, consider the choreography $p_1 \synchronous \{p_2\} \bullet p_1 \synchronous \{p_3\}$. Its corresponding distributed implementation would require to create two interactions $(p_1, \{p_2\})$ and $(p_1, \{p_3\})$. As such, the component that corresponds to $p_1$ ($B_1$) needs to interact $B_2$ and $B_3$ to know which interaction must be executed (depending on their current enable ports). However, if we create a copy of the ports, each port will be connected to one and only interaction, hence component $B_1$ can locally decide, without interacting with other components, on the interaction to be executed. 
As for correctness, consider the choreography $p_1 \synchronous \{p_2, p_3\} \bullet p_1 \synchronous \{p_2\}$. According to the choreography semantics, we should first execute  $p_1 \synchronous \{p_2, p_3\}$ then $p_1 \synchronous \{p_2\}$. However, if component $B_3$ is still executing the function just before the transition labeled with $p_3$, $B_1$ would interact with $B_2$ and $B_3$ to know which interaction to execute. As $p_3$ is not currently enabled, it would have executed the interaction connected with $p_2$ only, hence violating the sequential semantics. 
\end{remark}

\newcommand{\semfct}[1]{\llbracket #1 \rrbracket}
\newcommand{\rmnew}[0]{\mathrm{new}}
\begin{definition}[Send/Receive] 
\label{def:send-receive}

\[
\semfct{\itpsas [g,f] \synchronous \{ \rcvstt \} } (\gamma(B_1, \ldots, B_n), \context) = (\gamma'(B_1', \ldots, B_n'), \context'), \text{with:}
\]

\begin{itemize}
	\item
		$
		B'_k = 
		 \left\{ \begin{array}{ll}
		   (P_k, L'_k, T'_k) & \text{ if } k \in \stC(\itpsas [g, f]) \cup \stC(\rcvstt) \\
			B_k & \text{otherwise}
			\end{array}
			\right.
		$, where:
		\begin{itemize}
		\item
		$L_k' = L_k \cup \{ l_k^{\rm new} \}$
		\item $T'_k = T_k \, \cup \, \left\{ \begin{array}{ll}
		   \{\context(B_k) \synch[\copyport(\itpsas), g, f] l^{\rm new}_k\} & \text{ if } \itpsas [g, f] \in B_k.{\cal P}^{\rm ss} \cup  B_k.{\cal P}^{\rm as}\\
			 \{\context(B_k) \synch[\copyport(p_k), \true, p_k.\ufct] l^{\rm new}_k\} & \text{ if } p_k \in \rcvstt
			\end{array}
			\right.
			$
		\end{itemize}
	\item
	$\gamma' = \gamma \cup \{ (\copyport(\itpsas), \{\copyport(p_i) \mid p_i \in \rcvstt  \} ) \}$,
	\item
	$
	\context'(B'_k) =
	\left\{
	\begin{array}{ll}
		l^{\rm new}_k & \text{if } k \in \stC(\itpsas [g, f]) \cup \stC( \rcvstt ) \\
			\context(B_k) & \text{otherwise}
	\end{array}
	\right.
	$.
\end{itemize}
\end{definition}
Atomics components that do not participate in the send/receive choreography remain unchanged.
Atomic components that participate in the send/receive are updated by adding a transition from their context location to a new location ($l^{\rm new}_{k}$).
We label this transition with a copy of the corresponding port.
We create an interaction that connects the send ports to the receive ports.
The new context becomes the new created location.
\begin{example}[Send/Receive]
Figure~\ref{example:send-receive} shows an abstract example on how to transform a simple send/receive choreography, $\exchannel{b1}{S} \synch \{\exchannel{b2}{R},\; \exchannel{b3}{R}\}$, into an initial system consisting of three components with interfaces: $\exchannel{b1}{S}$ (send, synchronous or asynchronous), $\exchannel{b2}{R}$ (receive), and $\exchannel{b3}{R}$ (receive), respectively. 
\begin{figure}[t]
    \centering
    \begin{tikzpicture}[bip]
        \component{B1}{minimum size=1cm}{}{
        	\node[place]  (c1l1)  {$q^1_1$};
			\node[place]  (c1l2)  [below of = c1l1]{$q^1_2$};   
			\path[->] (c1l1) edge [] node [label=right:{\footnotesize{$\exchannel{b1}{S}$}}]   {} (c1l2);   		
            
        };
        \node[exportsend] (c1l2) at ($(B1.north)$) [label=below:$\exchannel{b1}{S}$]    {};

       \component{B2}{minimum size=1cm,  right=0.5cm of B1}{}{
            \node[place]  (c2l1)   {$q^2_1$};
            \node[place]  (c2l2) [below of = c2l1] {$q^2_2$};
            \path[->] (c2l1) edge [] node [label=right:{\footnotesize{$\exchannel{b2}{R}$}}]   {} (c2l2);
        };
        \node[exportreceive] (c2l2) at ($(B2.north)$) [label=below:$\exchannel{b2}{R}$]    {};
        
       \component{B3}{minimum size=1cm, right=0.5cm of B2}{}{
            \node[place]  (c3l1)   {$q^3_1$};
            
            \node[place]  (c3l2) [below of = c3l1] {$q^3_2$};
            \path[->] (c3l1) edge [] node [label=right:{\footnotesize{$\exchannel{b3}{R}$}}]   {} (c3l2);            
            
        };
        \node[exportreceive] (c3l2) at ($(B3.north)$) [label=below:$\exchannel{b3}{R}$]    {};
               
        \draw[-] (c1l2) -- ++(0,5mm) -| (c2l2) -- ++(0,5mm) -| (c3l2);

    \end{tikzpicture}
    \caption{Send/Receive Transformation}
        \label{example:send-receive}
\end{figure}
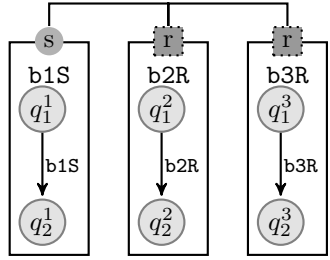
\end{example}
%
\subsubsection{Branching Composition}
%
Recall that conditional master branching of the form $B_i \oplus \{p^l_i[g_i,f_i]:\st_l\}_{l \in L}$, allows for the modeling of conditional choice between several choreographies.
The choice is made by a specific component ($B_i$), which depending on its internal state would enable some its guards ($g_i$). 
Accordingly, it  notifies the appropriate components by sending  a label ($p^l_i$), to follow the taken choice (i.e., the corresponding choreography, $\st_l$).
We apply branching by independently integrating the choreography for each choice.
This can be done by letting $B_i$ notifying the participants, i.e., 
 $\stC(B_i \oplus \{p^l_i[-]:\st_l\}_{l \in L} ) \setminus \{i\}$,
 of the choreography ($\st_l$) of that choice ($p^l_i$). For that purpose, we create new receive ports ($\{p^{\controlrcv_l}_{k}\}_{k \in K}$) to be able to receive the corresponding choice.

For this, we define a union operator, noted $\unionst$, that takes a set of systems with their contexts and (1) unions all of their locations, transitions and ports; then (2) updates the contexts of the obtained components by joining each of their input contexts with internal transitions.
Therefore, after applying branching we guarantee that each component will have one and only one context location.
Formally, operator $\unionst$ is defined as follows. 
\begin{definition}[Union]
The union of systems with their contexts $\{(S_l, \context_l)\}_{l \in L}$, where $S_l = \gamma^l(B^l_1, \ldots, B^l_n)$ and $B_i^l = (P_i^l, X_i, L_i^l, T_i^l)$ for $i \in [1,n]$ and $l \in L$, noted $\unionst(\{(S_l, \context_l)\}_{l \in L}) $, is defined as the system with context $\left( \gamma(B_1, \ldots, B_n), \context \right)$, where:
\begin{itemize}
\item $\gamma = \bigcup_{l \in L} \gamma^l$;
\item $B_i = (\bigcup_{l \in L} P_i^l, \bigcup_{l \in L} X_i^l, \bigcup_{l \in L} L_i^l \cup \{l_i^u\}_{l \in L}, \bigcup_{l \in L} T_i^l\, \cup\, T_i^{\mathtt{merge}})$ with $l_i^u$ a new location and $T_i^{\mathtt{merge}} = \{\context_l(B_i^l) \synch[\epsilon] q^c_i \mid \l  \in L\}$;
\item $\context(B_i) = l_i^u$ for $i \in [1,n]$.
\end{itemize}
\end{definition}
Then, branching as described by independently applying each choice, then doing the union.
\begin{definition}[Branching] 
\label{def:branching}
\[
\begin{array}{l}
\semantics[B_i \oplus \{p^l_i{[g_l,f_l]}:\st_l\}_{l \in L}](S, \context) \\
\quad = \unionst \left(\{ \semfct{\st_l}\semfct{p^l_i{[g_l,f_l]} \synchronous \{p^{\controlrcv_l}_{k}[\emptyset]\}_{k \in K}} (S, \context) \}_{l \in L} \right) 
\end{array}
\]
Where, $K = \stC(B_i \oplus \{p^l_i[-]:\st_l\}_{l \in L} ) \setminus \{i\}$. 
\end{definition}
\begin{remark}
Note that we require to notify all the participants of a choice and not only the start components.
Consider the following choreography (where $\alpha$ and $\beta$ denote some choreographies):
\[
B_1 \oplus \{p^l_1[-]: p_2[-] \synchronous p_3[-] \bullet \alpha;\; p^l_2[-]: p_2[-] \synchronous p_3[-] \bullet \beta\}
\]
In this choreography, if we would have not sent the choice made by component $1$ to component $3$, then component $3$ cannot know about the decision that was taken by component $1$.
Hence, it cannot decide whether to follow choreography $\alpha$ or $\beta$ afterwards.
\end{remark}
\begin{example}[Branching]
Figure~\ref{example:branching} shows an abstract example on how to apply a branching operation that consists of two choices  $B_1 \oplus \{b1^{l_1}[g_1,f_1]: \st_1, b2^{l_2}[g_2,f_2]:\st_2\}$.
First, we add choice transitions to component $B_1$ and synchronize them with the participants of $\st_1$ and $\st_2$, e.g., $B_2$ and $B_3$.
Then, we apply the choreographies accordingly.
Finally, we merge the contexts with internal transitions. 

\begin{figure}[tb]
    \centering
    \begin{tikzpicture}[bip]
        \component{B1}{minimum size=1.5cm}{}{
        	\node[place]  (c1l1)  {$q^1_1$};
			\node[place]  (c1l2)  [below of = c1l1,xshift=-0.75cm]{$q^1_2$};   
			\node[place]  (c1l3)  [below of = c1l1,xshift=0.75cm]{$q^1_3$};   
			\path[->] (c1l1) edge [] node [align=left, left, yshift=0.15cm]   {\footnotesize{$g_1$}\\\footnotesize{$\exchannel{b1}{}^{l_1}$} \\ \footnotesize{$f_1$}} (c1l2);
			\path[->] (c1l1) edge [] node [align=right, right, yshift=0.15cm]   {\footnotesize{$g_2$}\\ \footnotesize{$\exchannel{b1}{}^{l_2}$} \\ \footnotesize{$f_2$}} (c1l3);
			
			\node[place]  (c1l4)  [below of = c1l2]{$q^1_4$};   
			\node[place]  (c1l5)  [below of = c1l3]{$q^1_5$};   
			
			\path[dashed,->] (c1l2) edge [] node [label=right:{\footnotesize{$\st_1$}}]   {} (c1l4);
			\path[dashed,->] (c1l3) edge [] node [label=right:{\footnotesize{$\st_2$}}]   {} (c1l5);
			
			\node[place]  (c1l6)  [below of = c1l4,xshift=0.75cm]{$q^1_6$};   
			\path[->] (c1l4) edge [] node [label=right:{\footnotesize{$\varepsilon$}}]   {} (c1l6);
			\path[->] (c1l5) edge [] node [label=right:{\footnotesize{$\varepsilon$}}]   {} (c1l6);
        };
        \node[exportssend] (c1l2) at ($(B1.north west)!.3!(B1.north east)$) [label=below:$\exchannel{b1}{}^{l_1}$]    {};
        \node[exportssend] (c1l3) at ($(B1.north west)!.7!(B1.north east)$) [label=below:$\exchannel{b1}{}^{l_2}$]    {};

       \component{B2}{minimum size=1.5cm,  right=0.25cm of B1}{}{
        	\node[place]  (c2l1)  {$q^2_1$};
			\node[place]  (c2l2)  [below of = c2l1,xshift=-0.75cm]{$q^2_2$};   
			\node[place]  (c2l3)  [below of = c2l1,xshift=0.75cm]{$q^2_3$};   
			\path[->] (c2l1) edge [] node [align=left,left, yshift=0.15cm]   {\footnotesize{$\exchannel{b2}{}^{cr_1}$}} (c2l2);
			\path[->] (c2l1) edge [] node [align=right, right, yshift=0.15cm]   {\footnotesize{$\exchannel{b2}{}^{cr_2}$}} (c2l3);
			
			\node[place]  (c2l4)  [below of = c2l2]{$q^2_4$};   
			\node[place]  (c2l5)  [below of = c2l3]{$q^2_5$};   
			
			\path[dashed,->] (c2l2) edge [] node [label=right:{\footnotesize{$\st_1$}}]   {} (c2l4);
			\path[dashed,->] (c2l3) edge [] node [label=right:{\footnotesize{$\st_2$}}]   {} (c2l5);
			
			\node[place]  (c2l6)  [below of = c2l4,xshift=0.75cm]{$q^2_6$};   
			\path[->] (c2l4) edge [] node [label=right:{\footnotesize{$\varepsilon$}}]   {} (c2l6);
			\path[->] (c2l5) edge [] node [label=right:{\footnotesize{$\varepsilon$}}]   {} (c2l6);
        };
        \node[exportreceive] (c2l2) at ($(B2.north west)!.3!(B2.north east)$) [label=below:$\exchannel{b2}{}^{cr_1}$]    {};
        \node[exportreceive] (c2l3) at ($(B2.north west)!.7!(B2.north east)$) [label=below:$\exchannel{b2}{}^{cr_2}$]    {};

       \component{B3}{minimum size=1.5cm, right=0.25cm of B2}{}{
        	\node[place]  (c3l1)  {$q^3_1$};
			\node[place]  (c3l2)  [below of = c3l1,xshift=-0.75cm]{$q^3_2$};   
			\node[place]  (c3l3)  [below of = c3l1,xshift=0.75cm]{$q^3_3$};   
			\path[->] (c3l1) edge [] node [align=left, left, yshift=0.15cm]   {\footnotesize{$\exchannel{b3}{}^{cr_1}$}} (c3l2);
			\path[->] (c3l1) edge [] node [align=right, right, yshift=0.15cm]   {\footnotesize{$\exchannel{b3}{}^{cr_2}$}} (c3l3);
			
			\node[place]  (c3l4)  [below of = c3l2]{$q^3_4$};   
			\node[place]  (c3l5)  [below of = c3l3]{$q^3_5$};   
			
			\path[dashed,->] (c3l2) edge [] node [label=right:{\footnotesize{$\st_1$}}]   {} (c3l4);
			\path[dashed,->] (c3l3) edge [] node [label=right:{\footnotesize{$\st_2$}}]   {} (c3l5);
			
			\node[place]  (c3l6)  [below of = c3l4,xshift=0.75cm]{$q^3_6$};   
			\path[->] (c3l4) edge [] node [label=right:{\footnotesize{$\varepsilon$}}]   {} (c3l6);
			\path[->] (c3l5) edge [] node [label=right:{\footnotesize{$\varepsilon$}}]   {} (c3l6);         
        };      
        \node[exportreceive] (c3l2) at ($(B3.north west)!.3!(B3.north east)$) [label=below:$\exchannel{b3}{}^{cr_1}$]    {};
        \node[exportreceive] (c3l3) at ($(B3.north west)!.7!(B3.north east)$) [label=below:$\exchannel{b3}{}^{cr_2}$]    {};

        \draw[-] (c1l2) -- ++(0,9mm) -| (c2l2) -- ++(0,9mm) -| (c3l2);
        \draw[-] (c1l3) -- ++(0,5mm) -| (c2l3) -- ++(0,5mm) -| (c3l3);

    \end{tikzpicture}
        \caption{Branching transformation}
        \label{example:branching}
\end{figure}
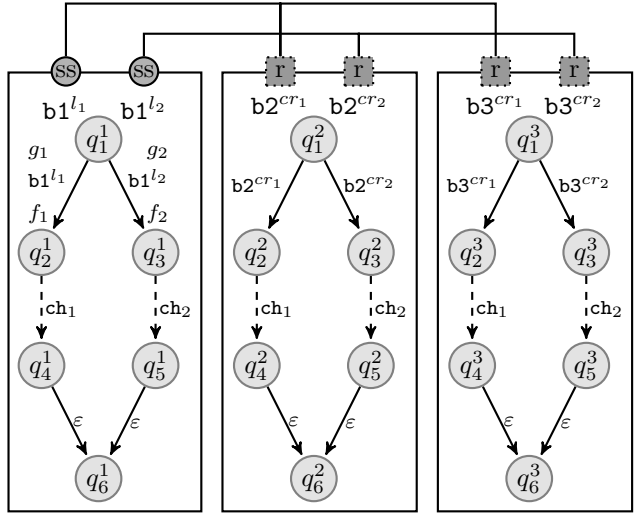
\end{example}
%
\subsubsection{Loop Composition}
%
Loop $\while(\sndit[g,f]) \{\st\}$, allows for the modeling of a conditional repeated choreograph $\st$. The condition is evaluated by a specific component, which will notify, through the port $p_i$, the participants of the choreography to either re-execute it or break.
\begin{definition}[Loop] 
\label{def:loop}
\[
\begin{array}{l}
\text{let } K = \stC(\st) \setminus \{i\} \\
\text{let } (\gamma^{\mathtt{t}}\left( B_1^{\mathtt{t}}, \ldots, B_n^{\mathtt{t}} \right), \context^{\mathtt{t}}) = \semfct{\st} \semfct{ \sndit [g,f] \synchronous \{ \prit_k^{\tt cont}[\emptyset] \}_{k \in K} }(S, \context) \\
\text{let } (P_i^{\mathtt{t}}, -, L_i^{\mathtt{t}}, T_i^{\mathtt{t}}) = B_i^{\mathtt{t}}, \text{ for } i \in [1, n]\\
\text{in } \semfct{ \ttwhile(\sndit[g,f]) \st\ \ttend }(S, \context) = (\gamma' \left( B_1', \ldots, B_n'\right) , \context') \\
\text{where: }\\
\text{let } p^\mathtt{f}_j \text{and } l^c_j \text{ be new synchronous ports and locations, for } j \in K \cup \{i\}
\end{array}
\]
\begin{itemize}
	\item
	$
	P'_j = P^{\mathtt{t}}_j \cup
	\left\{
	\begin{array}{ll}
		\{ p^\mathtt{f}_j \} & \text{if } j \in K \cup \{i\} \\
		\emptyset & \text{otherwise}
	\end{array}
	\right.
	$;
	\item $L'_j = L_j^{\mathtt{t}} \cup
	\left\{
	\begin{array}{ll}
		\{ l^c_j \} & \text{if } j \in K \cup \{i\} \\
		\emptyset & \text{otherwise}
	\end{array}
	\right.
	$;
	\item
	$
	T_j' = T_j^{\mathtt{t}} \cup
	\left\{
	\begin{array}{ll}
	\{ \context^{\mathtt{t}}(B_j) \synch[\epsilon]  \context(B_j), \context(B_j) \synch[p^\mathtt{f}_j, \true, \emptyset]  l^c_j \} & \text{if } j = i \\
	\{ \context^{\mathtt{t}}(B_j) \synch[\epsilon]  \context(B_j), \context(B_j) \synch[p^\mathtt{f}_j, \lnot g, \emptyset]  l^c_j \} & \text{if } j \in K \setminus \{i\} \\
	\emptyset & \text{otherwise}
	\end{array}
	\right.
	$;
	\item $\gamma' = \gamma^{\mathtt{t}} \cup \{ (p^\mathtt{f}_i, \{p^\mathtt{f}_j\}_{j \in K}) \} $;
	\item
	$
	\context'(B'_j) =
	\left\{
	\begin{array}{ll}
	l_j^c & \text{if } j \in  K \cup \{i\} \\
	\context(B_j) & \text{otherwise}
	\end{array}
	\right.
	$.
\end{itemize}
\end{definition}
Transitions are updated by adding the reset and loop transitions.
The condition is evaluated by a specific component, which will notify, through the port $p_i$, the participants of the choreography to either re-execute it or break.
The context is updated to be the location associated with the end of the loop. 
\begin{example}[Loop]
Figure~\ref{example:loop} shows an example of application of a loop operation guided by component $B_1$ and where the participants are components $B_1$, $B_2$ and $B_3$. 

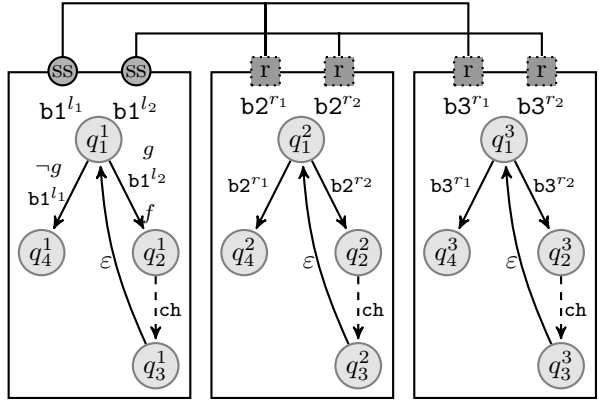
\begin{figure}[t]
    \centering
    \begin{tikzpicture}[bip]
        \component{B1}{minimum size=1cm}{}{
        	\node[place]  (c1l1)  {$q^1_1$};
			\node[place]  (c1l2)  [below of = c1l1,xshift=0.75cm]{$q^1_2$};   
			\node[place]  (c1l4)  [below of = c1l1,xshift=-0.75cm]{$q^1_4$};   
			\path[->] (c1l1) edge [] node [align=center,right, yshift=0.15cm]   {\footnotesize{$g$}\\ \footnotesize{$\exchannel{b1}{}^{l_2}$} \\ \footnotesize{$f$}} (c1l2);
			\path[->] (c1l1) edge [] node [align=center, left, yshift=0.15cm]   {\footnotesize{$\lnot g$} \\ \footnotesize{$\exchannel{b1}{}^{l_1}$}} (c1l4);
			
			\node[place]  (c1l3)  [below of = c1l2]{$q^1_3$};   
			
			\path[dashed,->] (c1l2) edge [] node [label=right:{\footnotesize{$\st$}}]   {} (c1l3);
			\draw[bend left=10,->]  (c1l3) to node [auto] {$\varepsilon$} (c1l1);
        };
        \node[exportssend] (c1l2) at ($(B1.north west)!.3!(B1.north east)$) [label=below:$\exchannel{b1}{}^{l_1}$]    {};
        \node[exportssend] (c1l3) at ($(B1.north west)!.7!(B1.north east)$) [label=below:$\exchannel{b1}{}^{l_2}$]    {};

       \component{B2}{minimum size=1cm,  right=.25cm of B1}{}{
        	\node[place]  (c2l1)  {$q^2_1$};
			\node[place]  (c2l2)  [below of = c2l1,xshift=0.75cm]{$q^2_2$};   
			\node[place]  (c2l4)  [below of = c2l1,xshift=-0.75cm]{$q^2_4$};   
			\path[->] (c2l1) edge [] node [align=center,right, yshift=0.15cm]   {\footnotesize{$\exchannel{b2}{}^{r_2}$}} (c2l2);
			\path[->] (c2l1) edge [] node [align=center, left, yshift=0.15cm]   {\footnotesize{$\exchannel{b2}{}^{r_1}$}} (c2l4);
			
			\node[place]  (c2l3)  [below of = c2l2]{$q^2_3$};   
			
			\path[dashed,->] (c2l2) edge [] node [label=right:{\footnotesize{$\st$}}]   {} (c2l3);
			\draw[bend left=10,->]  (c2l3) to node [auto] {$\varepsilon$} (c2l1);
        };
        \node[exportreceive] (c2l2) at ($(B2.north west)!.3!(B2.north east)$) [label=below:$\exchannel{b2}{}^{r_1}$]    {};
        \node[exportreceive] (c2l3) at ($(B2.north west)!.7!(B2.north east)$) [label=below:$\exchannel{b2}{}^{r_2}$]    {};

       \component{B3}{minimum size=1cm, right=0.25cm of B2}{}{
        	\node[place]  (c3l1)  {$q^3_1$};
			\node[place]  (c3l2)  [below of = c3l1,xshift=0.75cm]{$q^3_2$};   
			\node[place]  (c3l4)  [below of = c3l1,xshift=-0.75cm]{$q^3_4$};   
			\path[->] (c3l1) edge [] node [align=center,right, yshift=0.15cm]   {\footnotesize{$\exchannel{b3}{}^{r_2}$}} (c3l2);
			\path[->] (c3l1) edge [] node [align=center,left, yshift=0.15cm]   {\footnotesize{$\exchannel{b3}{}^{r_1}$}} (c3l4);
			
			\node[place]  (c3l3)  [below of = c3l2]{$q^3_3$};   
			
			\path[dashed,->] (c3l2) edge [] node [label=right:{\footnotesize{$\st$}}]   {} (c3l3);
			\draw[bend left=10,->]  (c3l3) to node [auto] {$\varepsilon$} (c3l1);      
        };      
        \node[exportreceive] (c3l2) at ($(B3.north west)!.3!(B3.north east)$) [label=below:$\exchannel{b3}{}^{r_1}$]    {};
        \node[exportreceive] (c3l3) at ($(B3.north west)!.7!(B3.north east)$) [label=below:$\exchannel{b3}{}^{r_2}$]    {};
        
        \draw[-] (c1l2) -- ++(0,9mm) -| (c2l2) -- ++(0,9mm) -| (c3l2);
        \draw[-] (c1l3) -- ++(0,5mm) -| (c2l3) -- ++(0,5mm) -| (c3l3);

    \end{tikzpicture}
        \caption{Loop composition transformation}
        \label{example:loop}
\end{figure}
\end{example}
%
\subsubsection{Sequential Composition}
%
The binary operator $\bullet$ allows to sequentially compose two choreographies, $\st_1 \bullet \st_2$. 
For this, its semantics is defined by (1) applying $\st_1$; (2) notifying the start of $\st_2$; and finally (3) applying $\st_2$.  As we require that $\st_1$ must terminate before the start of $\st_2$, we need to synchronize all the end components of $\st_1$ with all the start components of $\st_2$.
To do so, it is sufficient to pick one of the end components of $\st_1$ and create a synchronous send port, which is connected to new receive ports added to the remaining end components of $\st_1$ and start components of $\st_2$.
Moreover, the application of the sequential composition guarantees that each component of the resulting system consists of exactly one state, provided that the context of each component of the initial system consists of one state.
Formally, the semantics of the sequential composition is defined as follows. 
\begin{definition}[Sequential Composition] 
\label{def:sequential}
\[\semantics[\st_1 \bullet \st_2](S, \context) = 
\semantics[\st_2]\semantics[\st_{synch}]\semantics[\st_1](S, \context), \text{with:}
\]
$\st_{synch} = p_i^{\controlsend}[\true, \emptyset] \synchronous \{p^{\controlrcv}_j[\true, \emptyset]\}_{j \in J}$ such that: (1) $i \in \stend(\st_1)$; (2) $J = \stend(\st_1) \cup  \ststart(\st_2) \setminus \{i\}$; (3) $p_i^{\controlsend}$ is a new synchronous send port to be added to ${\cal P}_i^{ss}$; and (4) $\{p_j^{\controlrcv}\}_{j \in J}$ are new receive ports to be added to ${\cal P}_j^{r}$.
\end{definition}
\begin{example}[Sequential composition]
Figure~\ref{example:sequential} shows an abstract example on how to transform sequential composition of two choreographies, $\st_1 \bullet \st_2$, into an initial system consisting of five components. Here we only consider components that are involved in those choreographies, where (1) components $b_1$, $b_2$, $b_3$ and $b_4$ are involved in choreography $\st_1$; and (2) components $b_1$, $b_2$, $b_3$ and $b_5$ are involved in choreography $\st_2$. Note, components that are not involved are kept unchanged.
The transformation requires to: (1) apply first choreography $\st_1$ to its participated components (i.e., $b_1$, $b_2$, $b_3$ and $b_4$); (2) synchronize the end of choreography $\st_1$ (e.g., $b_1$) with the start of choreography $\st_2$ (e.g., $b_2$ and $b_3$). To do so, we create a synchronous send port to one of the end components of $\st_1$ (e.g., $b_1^{cs}$) and connect it to all the remaining end components of $\st_1$ (e.g., $\emptyset$ and the start components of $\st_2$ (e.g., $b_2$ and $b_3$); finally (3) we apply choreography $\st_2$.
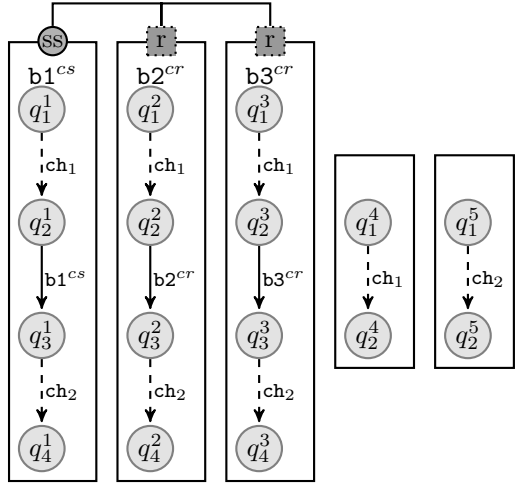
\begin{figure}[t]
    \centering
    \begin{tikzpicture}[bip]
        \component{B1}{minimum size=1cm}{}{
        	\node[place]  (c1l1)  {$q^1_1$};
			\node[place]  (c1l2)  [below of = c1l1]{$q^1_2$};   
			\path[dashed, ->] (c1l1) edge [] node [label=right:{\footnotesize{$\st_1$}}]   {} (c1l2);
			\node[place]  (c1l3)  [below of = c1l2]{$q^1_3$}; 
			\path[->] (c1l2) edge [] node [label=right:{\footnotesize{$\exchannel{b1}{}^{cs}$}}]   {} (c1l3);
			
			\node[place]  (c1l4)  [below of = c1l3]{$q^1_4$};   
			\path[dashed, ->] (c1l3) edge [] node [label=right:{\footnotesize{$\st_2$}}]   {} (c1l4);
            
        };
        \node[exportssend] (c1l3) at ($(B1.north)$) [label=below:$\exchannel{b1}{}^{cs}$]    {};

       \component{B2}{minimum size=1cm,  right=0.25cm of B1}{}{
            \node[place]  (c2l1)   {$q^2_1$};
            \node[place]  (c2l2) [below of = c2l1] {$q^2_2$};
            \path[dashed, ->] (c2l1) edge [] node [label=right:{\footnotesize{$\st_1$}}]   {} (c2l2);
            
			\node[place]  (c2l3)  [below of = c2l2]{$q^2_3$}; 
			\path[->] (c2l2) edge [] node [label=right:{\footnotesize{$\exchannel{b2}{}^{cr}$}}]   {} (c2l3);
			
			\node[place]  (c2l4)  [below of = c2l3]{$q^2_4$};   
			\path[dashed, ->] (c2l3) edge [] node [label=right:{\footnotesize{$\st_2$}}]   {} (c2l4);            
              
        };
        \node[exportreceive] (c2l3) at ($(B2.north)$) [label=below:$\exchannel{b2}{}^{cr}$]    {};

       \component{B3}{minimum size=1cm, right=0.25cm of B2}{}{
  \node[place]  (c3l1)   {$q^3_1$};
            \node[place]  (c3l2) [below of = c3l1] {$q^3_2$};
            \path[dashed, ->] (c3l1) edge [] node [label=right:{\footnotesize{$\st_1$}}]   {} (c3l2);
            
			\node[place]  (c3l3)  [below of = c3l2]{$q^3_3$}; 
			\path[->] (c3l2) edge [] node [label=right:{\footnotesize{$\exchannel{b3}{}^{cr}$}}]   {} (c3l3);
			
			\node[place]  (c3l4)  [below of = c3l3]{$q^3_4$};   
			\path[dashed, ->] (c3l3) edge [] node [label=right:{\footnotesize{$\st_2$}}]   {} (c3l4);                   
        };
                \node[exportreceive] (c3l4) at ($(B3.north)$) [label=below:$\exchannel{b3}{}^{cr}$]    {};

       \component{B4}{minimum size=1cm, right=0.25cm of B3}{}{
            \node[place]  (c4l1)   {$q^4_1$};
            
            \node[place]  (c4l2) [below of = c4l1] {$q^4_2$};
            \path[dashed, ->] (c4l1) edge [] node [label=right:{\footnotesize{$\st_1$}}]   {} (c4l2);              
            
        };

       \component{B5}{minimum size=1cm, right=0.25cm of B4}{}{
            \node[place]  (c5l1)   {$q^5_1$};
            
            \node[place]  (c5l2) [below of = c5l1] {$q^5_2$};
            \path[dashed, ->] (c5l1) edge [] node [label=right:{\footnotesize{$\st_2$}}]   {} (c5l2);              
            
        };

        \draw[-] (c1l3) -- ++(0,5mm) -| (c2l3) -- ++(0,5mm) -| (c3l4);

    \end{tikzpicture}
    \caption{Sequential composition transformation}
        \label{example:sequential}
\end{figure}
\end{example}
%
\subsubsection{Parallel Composition}
%
The binary operator $\parallel$ allows for the parallel compositions of two independent choreographies. Two choreographies are independent if their participating components are disjoint. 
\begin{definition}[Independent Choreographies] 
Two choreographies $\st_1$ and $\st_2$ are said to be independent iff $\stC(\st_1) \cap \stC(\st_2) = \emptyset$.
\end{definition}
We consider independent choreographies to avoid conflicts and interleaving of executions within components.
In addition, this simplifies reasoning and writing choreographies as well as for efficient code generation.
Note that parallelizing independent choreographies implies that each component has a single execution flow. In case we have overlap, e.g.,
$p_1 \synchronous \{p_2, p_3\} \parallel p_1 \synchronous \{p_5\}$, we could split $p_1$ into two different components. Moreover, it is possible to enforce any arbitrary order of execution.  Further, we discuss other possible alternatives for handling this case. 
This would not reduce the expressiveness of our model as parallel execution flows can be modelled in separate components. 
The semantics of the parallel composition $\st_1 \parallel \st_2$ is simply defined by applying  $\st_1$ and $\st_2$ in any order, which leads to the same system as the two choreographies are independent, i.e., they behave on different set of components.
Moreover, the application of the parallel composition guarantees that each component of the resulting system consists of exactly one state, provided that the context of each component of the initial system consists of one state. 
\begin{definition}[Parallel Composition] 
\label{def:parallel}
\[\semantics[\st_1 \parallel \st_2](S, \context) = \semantics[\st_2]\semantics[\st_1](S, \context)
\]
\end{definition}
\begin{example}[Parallel Composition]
Figure~\ref{example:sequential} shows an abstract example on how to transform parallel composition of two choreographies, $\st_1 \parallel \st_2$, into an initial system consisting of five components.
Here, we consider that $\st_1$ (resp. $\st_2$) involves components $B_1$ and $B_2$ (resp. $B_3$ and $B_4$). 

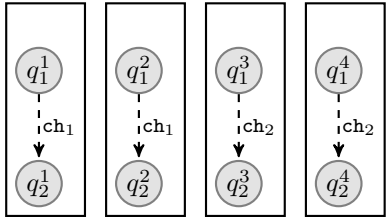
\begin{figure}[t]
    \centering
    \begin{tikzpicture}[bip]
        \component{B1}{minimum size=1cm}{}{
        	\node[place]  (c1l1)  {$q^1_1$};
			\node[place]  (c1l2)  [below of = c1l1]{$q^1_2$};   
			\path[dashed, ->] (c1l1) edge [] node [label=right:{\footnotesize{$\st_1$}}]   {} (c1l2);
        };

       \component{B2}{minimum size=1cm,  right=0.25cm of B1}{}{
            \node[place]  (c2l1)   {$q^2_1$};
            \node[place]  (c2l2) [below of = c2l1] {$q^2_2$};
            \path[dashed, ->] (c2l1) edge [] node [label=right:{\footnotesize{$\st_1$}}]   {} (c2l2);
        };

       \component{B3}{minimum size=1cm, right=0.25cm of B2}{}{
            \node[place]  (c3l1)   {$q^3_1$};
            \node[place]  (c3l2) [below of = c3l1] {$q^3_2$};
            \path[dashed, ->] (c3l1) edge [] node [label=right:{\footnotesize{$\st_2$}}]   {} (c3l2);           
        };
       \component{B4}{minimum size=1cm, right=0.25cm of B3}{}{
            \node[place]  (c4l1)   {$q^4_1$};
            \node[place]  (c4l2) [below of = c4l1] {$q^4_2$};
            \path[dashed, ->] (c4l1) edge [] node [label=right:{\footnotesize{$\st_2$}}]   {} (c4l2);           
        };

    \end{tikzpicture}
    \caption{Parallel composition transformation}
        \label{example:parallel}
\end{figure}
\end{example}
The following proposition is a straightforward consequence of the transformation associated with the $\parallel$ operator and the fact that the transformation of a choreography only modifies the component involved in this choreography.
\begin{proposition}
	If $\st_1$ and $\st_2$ are two independent choreographies, then $\semantics[\st_1 \parallel \st_2] = \semantics[\st_2 \parallel \st_1]$.
\end{proposition}
Consequently, synthesizing distributed systems for parallel choreographies can be done concurrently.
\begin{remark}
For parallelizing choreographies that have a component in common (i.e., not independent), we can still apply the parallel composition either by (1) enforcing any arbitrary order of execution.
As such, in the case of independent choreographies, true parallelism is achieved, otherwise, we apply them in any order to avoid non-deterministic execution; (2) using of product automata as defined in~\cite{TuostoG18}; (3) use of multiple execution flows (i.e., multi-threading within a component). 
\end{remark}
%

%
\section{Correctness of the synthesis method}
\label{sec:correctness}
%
In this section, we show the correctness of the transformations defined in the previous section.
More precisely, given a choreography $\st$, we show that the final configuration obtained by interpreting the choreography according to the choreography semantics is equivalent to the state obtained by transforming the choreography into a distributed system and executing the obtained system according to their semantics.
Both in choreographies and distributed systems, final configurations are states, that is mappings from the set of variables $X$ to the set of data values $\data$.
In choreographies and distributed systems, we consider an initial state where the variables are mapped to their default values. 
The proof is done by induction, following the syntax of choreographies.
Let us recall that the semantics of choreographies is in \figref{fig:semantics-choreographies} and the semantics of distributed systems is in \figref{fig:composite-rules}.
%
\subsection{Send/Receive}
%
We have to show that the state obtained after executing $\sndit \synchronous \{ \rcvstt \}$ is equivalent to the state obtained after executing $\semfct{\sndit \synchronous \{ \rcvstt \}}$.
We distinguish the case of synchronous send/receive from asynchronous one.
Note that in both cases, we create an interaction from the send to the receive port (we add the corresponding locations and transitions).
\begin{itemize}
	\item 
	In case the send port is synchronous.
	In the case of choreographies, the execution follows rule {\rulename{synch-sendrcv}} where the update functions are applied to the global state but they apply to disjoint variables.
	In the case of distributed systems, the execution follows rule \rulename{synch-send}.
	Each component applies its update function, independently.
	\item 
	In case the send port is synchronous.
	In the case of choreographies, the execution follows rules \rulename{asynch-sendrcv-1} and \rulename{asynch-sendrcv-2}.  	
	In the case of distributed systems, the execution follows rules \rulename{asynch-send} and \rulename{recv}.
	On the one hand, rule \rulename{asynch-send} corresponds to the application of \rulename{asynch-sendrcv-1}.
	On the other hand, rule \rulename{recv} corresponds to the application of \rulename{asynch-sendrcv-2}.
	Primitive $\sendrm$ in choreographies has the same effect has placing the value of the transferred variable in the buffer of the receiving components.
	Regarding the update functions, according to the distributed system semantics (atomic components):
	\begin{itemize}
	\item
	for the send component, the update function is applied after sending the transferred value of the variable of port $\sndit$ ($d$ in case of \rulename{asynch-send}).
	\item
	for the receive components, the update function is applied using the received value from the send ports.
	\end{itemize}
	Using the buffered communication, rule \rulename{asynch-send} is applied once, while rule \rulename{recv} is applied $|\{ \rcvstt \}|$ times.
	\end{itemize}
%
\subsection{Branching}
%
We have to show that the state obtained after executing $B_i \oplus \{ \{p^l_i[g_l,f_l]:\st_l\}_{l \in L} \}$ is equivalent to the state obtained after executing $\semfct{B_i \oplus \{ \{p^l_i[g_l,f_l]:\st_l\}_{l \in L} \}}$, provided that the transformation for each choreography ($\st_l$, $l \in L$) involved in the continuations is correct (induction hypothesis).
According to rule \rulename{master-branching}, whenever guard $g_l$ holds, update function $f_l$ is applied locally to the current state, and then $\st_l$ is applied.
According to Definition~\ref{def:branching}, we apply choreography $p^l_i{[g_l,f_l]} \synchronous \{p^{\controlrcv_l}_{k}[\emptyset]\}_{k \in K}$ to branch to the correct choreography according to the guard and to let component $B_i$ apply its update function.
Then, we apply choreography $\st_l$.
%
\subsection{Loops}
%
We have to show that the state obtained after executing $\ttwhile(\sndit[g,f]) \st\ \ttend$ is equivalent to the state obtained after executing $\semfct{\ttwhile(\sndit[g,f]) \st\ \ttend}$, provided that the transformation for choreography $\st$ is correct.
There are two rules for this case: \rulename{iterative-tt} and \rulename{iterative-ff}.
In Definition~\ref{def:loop}:
\begin{itemize}
\item	
rule \rulename{iterative-tt} corresponds to the transformation $\semfct{\sndit [g,f] \synchronous \{ \prit_k^{\tt cont}[\emptyset] \}_{k \in K}}$ and the iteration is ensured with added transition $\context^{\mathtt{t}}(B_j) \synch[\epsilon]  \context(B_j)$;
\item
rule \rulename{iterative-ff} corresponds to the added transition $\context(B_j) \synch[p^\mathtt{f}_j, \lnot g, \emptyset]  l^c_j $.
\end{itemize}
%
\subsection{Sequential Composition}
%
We have to show that the state obtained after executing $\st_1 \bullet \st_2$ is equivalent to the state obtained after applying $\semfct{\st_1 \bullet \st_2}$, provided that the transformations for $\st_1$ and $\st_2$ are correct (induction hypothesis).
According to the semantic rules of choreography, it boils down to proving that system $\semfct{\st_2}$ can start only when $\semfct{\st_1}$ has terminated.
This is guaranteed in Definition~\ref{def:sequential} by applying $\semfct{\st_{synch}}$ after applying $\semfct{\st_1}$.
The transformation $\semfct{\st_{synch}}$ synchronizes the end location of $\semfct{\st_1}$ with the start location of $\semfct{\st_2}$.
Since the context is reduced to a unique location, such synchronizations are well defined.
%
\subsection{Parallel Composition}
%
The case of parallel composition is similar to the case of sequential composition except that there is no need to add a synchronization after $\semfct{\st_1}$ as the two choreographies are independent.
\section{Code Generation}
\label{sec:code}
%
%
We describe the principle on how to generate distributed implementation from the generated components. 

Code generation takes as input a choreography and a configuration file containing the list of components with their corresponding interfaces/ports and variables.
Clearly, the choreography is defined with respect to components' ports, with  functions and guards defined with respect to components' variables. 
Code generation then automatically produces the corresponding implementation of each of the components. Following our transformation  into Distributed CBS in \secref{sec:tdcbs}, the obtained components have the following characteristics: (1) they do not have a location with outgoing send and receive ports; (2) a port is connected to exactly one interaction. As such, there is no conflicting interactions that can run concurrently.
Two interactions are said to be conflicting iff they share a common component.
Consequently, it is possible to generate fully distributed implementations, with no need for controllers (unlike~\cite{BonakdarpourBJQS12}) for managing multiparty interactions.
Hence, the number of exchanged messages will be divided by 2 for each execution of an interaction.

The code structure is depicted in Algorithm~\ref{algo:code-generation} that requires only send/receive primitives. We distinguish between two possible cases.
First, if all outgoing transitions are labeled with send ports, we pick a random enabled port, i.e., its guard evaluated to true. Then, we notify all the receive ports that are connected to the interaction containing that port.
If the port is a synchronous send port, the component waits for an acknowledgement from the corresponding receive components. Second, if all outgoing transitions are labeled with receive ports, the component waits until a message is ready/received in one of the receive ports. 
Upon receiving a message, we acknowledge its receipt if the port is connected to a synchronous interaction. 
Finally, we update the current state (update location and execute local function) of the component (\texttt{updateCurrentState()}) depending on the current outgoing transition.
It is worth mentioning that it is possible to provide a code generation w.r.t. a communication library (e.g., MPI, Java Message Service). In this case, the code generation can benefit from the features provided by the library, e.g., synchronous communication such as \texttt{MPI\_Ssend}.
\begin{algorithm}[ht]
 initialization()\;
 \While{true}{
    \If{all outgoing transitions are send}{
        port p = select enabled port, i.e., guard true\;
        notify all the receivers of the interaction that has port p\;
        \If{p is synchronous}{
           wait for ack. from the receivers\;
        }
   }
   \uElseIf{all outgoing transitions are receive}{
        wait until a message is ready in one of the outgoing receive ports\;
        port p = select message\;
         \If{interaction connected is synchronous}{
           send ack. to the corresponding send port\;
        }
   }
   updateCurrentState()\;
 }
 \caption{Pseudo-code - generated components.}
 \label{algo:code-generation}
\end{algorithm}
%
%
\section{Building Micro-Services Using Choreography}
%
%
Traditionally, distributed applications follow a monolithic architecture, i.e., all the services are embedded within the same application.
A new trend is to split complex applications up into smaller micro-services, where each micro-service can live on its own within a container.

We conduct a case study on a micro-service architecture to automatically derive the skeleton of each micro-service.
We use choreographies to describe the interactions between services.
The system consist of several communicating services to provide clients with system images.
Typical services include load balancing, authentication, fault-tolerance, installation, storage, configuration, and deployment.
The system also allows clients to request and install packages.

The corresponding global choreography $CH$ is defined in Listing~\ref{fig:global-choreography}. 
\begin{itemize}
\item $CH_1$:
A client ($\mathtt{c}$) sends a request to the \emph{gateway service} ($\mathtt{gs}$), which is the only visible micro-service to the client, containing the required version, revision, pool name, and an identifier to the testing data. $\mathtt{gs}$ forwards the request to the \emph{deploy environment service} ($\mathtt{des}$). 
$\mathtt{des}$ creates an environment id and returns it back to $\mathtt{gs}$, which in turn forwards it back to $\mathtt{c}$. 
\item $CH_2$:
$\mathtt{des}$ sends to the \emph{deploy application directory service} ($\mathtt{dads}$) and the \emph{deploy database service} ($\mathtt{dds}$) (i) required version, revision and pool name and (ii) testing data identifier and environment id, respectively. $\mathtt{c}$ keeps checking if the environment is ready, which is done through the gateway service with the help of the \emph{environment info. service} ($\mathtt{eis}$).
\item $CH_3$:
$\mathtt{dads}$ requests from the \emph{machine service} ($\mathtt{ms}$) and the \emph{setup service} ($\mathtt{ss}$) (i) a machine location from the pool and (ii) the package location, respectively. When $\mathtt{dads}$ receives the replies from both $\mathtt{ms}$ and $\mathtt{ss}$, it contacts the appropriate \emph{host machine} ($\mathtt{hm}_i$) by sending the package location. Then, $\mathtt{hm}_i$ sends its status to $\mathtt{des}$. $\mathtt{des}$ upon receiving the status update, it forwards it to the $\mathtt{eis}$. 
$\mathtt{dds}$ requests from the \emph{dumps service} ($\mathtt{dus}$) and the \emph{Database machines services} ($\mathtt{dms}$) (i) testing data location, and (ii) a database server, respectively. When $\mathtt{dds}$ receives the replies from both $\mathtt{dus}$ and $\mathtt{dbs}$, it contacts the appropriate \emph{database server} $\mathtt{hd}_j$ by sending the testing data location.
Then, $\mathtt{hd}_j$ sends its status to $\mathtt{des}$.
Upon receiving the status update, $\mathtt{des}$ forwards it to $\mathtt{eis}$. 
\end{itemize}
For each micro-service/component $\mathtt{m}$, we denote by $\mathtt{m}{SS}$, $\mathtt{m}{AS}$ $\mathtt{m}{R}$ a corresponding synchronous send, asynchronous send and receive port, respectively. 
\begin{lstlisting}[
	float=t,
	style=grammar,
	numbers=none,
	caption={Global choreography},
	label={fig:global-choreography}
	]
CH = CH$_1\;\bullet$ CH$_2 \, \bullet$ CH$_3$
CH$_1 = \exchannel{c}{SS} \synchsss \exchannel{gs}{R} \; \bullet \;  \exchannel{gs}{SS} \synchsss  \exchannel{des}{R} \; \bullet \; \exchannel{des}{AS} \synchsss \exchannel{gs}{R}$
CH$_2 =$ CH$_2^1\; \bullet$ CH$_2^2$
CH$_2^1 = \exchannel{gs}{SS} \synchsss  \exchannel{c}{R} \parallel (\exchannel{des}{AS}  \synchsss \exchannel{dads}{R} \; \bullet \; \exchannel{des}{AS}  \synchsss \exchannel{dads}{R})$ 
CH$_2^2 =$ while$(\exchannel{c}{SS}) \; \exchannel{c}{SS} \synchsss  \exchannel{gs}{R}\; \bullet$ 
	      $\exchannel{gs}{SS} \synchsss \exchannel{eis}{R} \; \bullet \exchannel{eis}{SS} \synchsss \exchannel{gs}{R}  \bullet \exchannel{gs}{SS}  \synchsss \exchannel{c}{R}$ end
CH$_3 = ($CH$_4\; \parallel$ CH$_5) \bullet$ CH$_6$
CH$_4 = $ CH$_4^1\; \bullet$ CH$_4^2 \;\bullet$ CH$_4^3$
CH$_4^1 = \exchannel{dads}{AS} \synchsss  \exchannel{ams}{R} \; \bullet \; \exchannel{dads}{AS} \synchsss  \exchannel{SS}{R}$
CH$_4^2 = \exchannel{ams}{SS} \synchsss  \exchannel{dads}{R} \; \parallel \; \exchannel{ss}{SS} \synchsss  \exchannel{dads}{R}$
CH$_4^3 = \mathtt{dads} \oplus \{l_i:  \exchannel{dads}{SS} \synchsss \exchannel{hm_i}{R} \; \bullet \; \exchannel{hm_i}{SS} \synchsss \exchannel{des}{R}\}$
CH$_5 = $ CH$_5^1\; \bullet$ CH$_5^2 \;\bullet$ CH$_5^3$
CH$_5^1 = \exchannel{dds}{AS} \synchsss  \exchannel{dus}{R} \; \bullet \; \exchannel{dds}{AS} \synchsss  \exchannel{SS}{R}$
CH$_5^2 = \exchannel{dus}{SS} \synchsss  \exchannel{dds}{R} \; \parallel \; \exchannel{dms}{SS} \synchsss  \exchannel{dads}{R}$
CH$_5^3 = \mathtt{dds} \oplus \{l_i:  \exchannel{dds}{SS} \synchsss \exchannel{hd_i}{R} \; \bullet \; \exchannel{hd_i}{SS} \synchsss \exchannel{des}{R}\}$
CH$_6 = \exchannel{des}{AS} \synchsss  \exchannel{eis}{R}$
\end{lstlisting}

Given the global choreography, we automatically synthesize the code of each component. Note that, in practice, the above choreography may be updated to full-fill new requirements by updating/adding/removing new micro-services.
This would require a drastic effort to re-implement the communication logic between components, which is tedious, error-prone and very time-consuming.
Using our method, we only require to update the global choreography, and then automatically generate the implementation of the components. 
\section{Transformation to Promela}
\label{sec:promela}
%
%
\paragraph{Overview}
Given a system $S = (B, \texttt{init})$, with $B = \gamma(B_1, \ldots, B_n)$,  produced by applying the set of transformations corresponding to a given choreography $\st$, we define a translation of $S$ into \texttt{Promela}~\cite{promelaspin}.
The \texttt{Promela} version of the system has the same behavior as $S$ but it can be verified with respect to properties specified in Linear Temporal Logic (LTL).

The transformation to \texttt{Promela} is realized mainly by two functions (1) \texttt{createChannels}, which generates global channels (in Promela) that are used to transfer messages between processes; (2) \texttt{createProcess}, which generates the code that corresponds to each of the components. We use the \texttt{append} call to add \texttt{Promela} code to the generated file.
Listing~\ref{lst:stopromela} depicts code generation for a system $S$ to \texttt{Promela}. 
\begin{lstlisting}[
	style=pseudoj,
	numbers=none,
	caption={Main Code Generation from System $S$ to \texttt{Promela}},
	label={lst:stopromela},
	float=t
	]
/*@\textbf{createPromela()}@*/ {
  createChannels();
  foreach $B_i$ {
    createProcess(i);
  }
}
\end{lstlisting}
\begin{lstlisting}[
	style=pseudoj,
	caption={\texttt{createChannels} Skeleton},
	label={lst:createchannel},
	float=t
	]
/*@\textbf{createChannels()}@*/ 
  foreach $a \in \gamma$, where $a =(p_s, \{p_r^i\}_{i \in I})$ {
    foreach  $p \in \{p_r^i\}_{i \in I}$ {
      if($\mathtt{isSSend}(p_s)$) 
        append chan channelP = [0] of {ps.dtype};
      else
        append chan channelP = [MAX_LEN] of {ps.dtype};
      end 
    end 
  end
\end{lstlisting}
\paragraph{Function \texttt{createChannels}}
The main skeleton of the \texttt{createChannels} is depicted in Listing~\ref{lst:createchannel}.
For every receive port, we create a channel (Promela's message carrier type). The type of the  channel is the data type of the corresponding send port (i.e., $p.\dtype$).
For synchronous (resp. asynchronous) ports, we use a channel of length 0 (resp. $\mathtt{MAX\_LEN}$).
\paragraph{Function \texttt{createProcess}}
The main skeleton of the \texttt{createProcess} is depicted in Listing~\ref{lst:createprocess}.
For every component $B_i$, we create a process in Promela containing: (1) a variable that will hold the current location of the component, which is initialized to the initial location of the component; a (2) the variables of the component; and (3) the code generated of the LTS implementation of the component.
\begin{lstlisting}[
	style=pseudoj,
	caption={\texttt{createProcess} Skeleton},
	label={lst:createprocess},
	float=t
	]
/*@\textbf{createProcess(int id)}@*/ {
  append proctype process(int id) {
    append int currentLocation = initialLocation;
    append currPort = _;
    append do
    append :: if
    append :: (all current outgoing trans. are send) ->
      append $p_s$ = pickEnablePort(); // w.r.t. guard
      append currPort = $p_s$;
      foreach  $p \in \{p_r^i\}_{i \in I}$, where $\exists a = (p_s, \{p_r^i\}_{i \in I}) \in \gamma$ {
        append channelP!(msg); 
      }
      append if
      append :: (all outgoing are synchronous send) ->
        foreach  $p \in \{p_r^i\}_{i \in I}$, where $\exists a = (p_s, \{p_r^i\}_{i \in I}) \in \gamma$ {
          append channelP?(_);
        }
      append fi;
      append ::  else -> // outgoing transitions are receive
        // listening to all current channels
        append if
         foreach p: currentLocation$\synch[p]$
            append ::(channelP?(val)) -> currPort = p;
            if(p is connected to synchronous send) {
               append channelP!(ack); 
            }
        append fi;
    append fi;
    // Update current location and execute location function
    // of the current outoing transition.
    append updateCurrentState(); 
  append  od;
  append }
}
\end{lstlisting}

\section{Case Study: Synthesizing a Correct Implementation of a Buying System}
\label{sec:case}
%
%
We consider a system consisting of four components: Buyer 1 ($B_1$), Buyer 2 ($B_2$), Seller ($S$) and Bank ($Bk$).
%
%
\subsection{Specification of the Buying System}
%
Buyer 1 sends a book title to the Seller, who replies to both buyers by quoting a price for the given book.
Depending on the price, Buyer 1 may try to haggle with Seller for a lower price, in which case Seller may either accept the new price or call off the transaction entirely.
At this point, Buyer 2 takes Seller's response and coordinates with Buyer 1 to determine how much each should pay.
In case Seller chose to abort, Buyer 2 would also abort.
Otherwise, it would keep negotiating with Buyer 1 to determine how much it should pay. Buyer 1, having a limited budget, consults with the bank before replying to Buyer 2.
Once Buyer 2 deems the amount to be satisfactory, he will ask the bank to pay the seller the agreed upon amount (Buyer 1 would be doing the same thing \emph{in parallel}).
%
\subsection{Synthesizing the Implementation}
\label{sec:example:synthesis}
%

\begin{lstlisting}[
	style=grammar,
	numbers=none,
	caption={Global choreography of the Buyer/Seller example},
	label={lst:buyerseller},
	float=t
	]
CH = B$_1$.S $\synchsss$ S.R$\ \bullet$ S.S $\synchsss \{$B$_1$.R, B$_2$.R$\} \ \bullet$ B$_1 \oplus \{$CH$_1$, $\epsilon \} \ \bullet$ CH$_2 \ \bullet$ CH$_7$ 
CH$_1$ = B$_1$.S $\synchsss$ S.R$\ \bullet$ S.S $\synchsss \{$B$_1$.R, B$_2$.R$\}$
CH$_2$ = B$_2 \oplus \{$CH$_3$, $\niltt\}$
CH$_3$ = while(B$_2$.C) $\{$
       B$_1$.C $\synchsss$ Bk.InfR$\ \bullet$ Bk.InfS $\synchsss$ B$_1$.R$\ \bullet$ B$_1$.C $\synchsss$ B$_2$.R
      $\} \ \bullet$ CH$_4$
CH$_4$ = CH$_5 \parallel$CH$_6$
CH$_5$ = B$_2$.MS $\synchsss$ Bk.MR$_2 \ \bullet$ Bk.MS$_2 \synchsss$ S.R
CH$_6$ = B$_1$.MS $\synchsss$ Bk.MR$_1 \ \bullet$ Bk.MS$_1 \synchsss$ S.R
CH$_7$ = B$_1$.E $\synchsss \niltt \parallel$ B$_2$.E $\synchsss \niltt \parallel $ Bk.E $\synchsss \niltt \parallel$ S.E $\synchsss \niltt$
\end{lstlisting}

\input{tikz/ltlEx}
\paragraph{Choreography}
We used the specification of the buying system to write a global choreography $\st$ that describe the expected interactions between the buyers and the seller.
The choreography is given Listing~\ref{lst:buyerseller}.
In the choreography, we prefix the names of the ports by the owning components.
Each port maps to a different functionality in the system so that, for example, \texttt{Bk.InfR} and \texttt{Bk.InfS} represent an interface for handling enquiries.
 \texttt{B$_i$.S} and \texttt{B$_i$.R} represent simple message send/receive interfaces for Buyer $i$ (similarly for \texttt{S.S} and \texttt{S.R}).
\paragraph{Synthesizing the distributed component-based system}
We apply our transformation to the choreography in Listing~\ref{lst:buyerseller} and obtain the distributed component-based system depicted in \figref{example:buyersellergenerated}.
The system consists of four components, one for each process involved in the choreography. Ports prefixed  with $\mathtt{cp}$ are controlled ports generated for synchronization following the transformations in Section~\ref{sec:transformation}. 
Interactions are used by the components to synchronize and communicate,  e.g., (1) $(B_1.S, \{S.R\})$, which allows buyer $B_1$ to request a quote from the seller; (2) $(B_2.cps_1, \{B_1.cpr_3, Bk.cpr_1, S.cpr_5\})$, which is used to broadcast the choice made by buyer $B_2$. In total, we generate $27$ interactions. 
Otherwise, the components evolve independently.
The components do not require controllers to execute; this ensures the efficiency of the implementation at runtime.
\paragraph{\texttt{Promela} version of the implementation}
To verify that the distributed implementation respects some desired properties, we apply our transformation of distributed component-based systems to \texttt{Promela} which constitutes a translation of the choreography behavior.

Because of the absence of procedures in \texttt{Promela}, we define the macros in Listing\ref{lst:macrosseller} for convenience and clarity.
 All of these macros accept a \texttt{Promela} channel (\texttt{ch}).
 We assume that \texttt{value} is a variable that contains the value that should be sent.
\begin{lstlisting}[
	style=pseudoProm,
	basicstyle=\small,
	numbers=none,
	caption={Promela Macros},
	label={lst:macrosseller},
	float=t
	]
#define recv(ch) ch?value
#define recvAck(ch) ch?(_)
#define send(ch) ch!value
#define sendAck(ch) ch!ack
#define synchRecv(ch) ch?value; sendAck(ch)
\end{lstlisting}

With the macros defined in Listing~\ref{lst:macrosseller}, the \texttt{Promela} code generated is depicted in Listing~\ref{lst:seller}.

\begin{lstlisting}[
	style=pseudoProm,
	basicstyle=\footnotesize,
	numbers=none,
	caption={Seller Process in Promela},
	label={lst:seller},
	float=t
	]
proctype Seller() {
  int currentLocation = $q_1$;
  currPort = _;
  int value;
  do
  ::if
    ::(currentLocation == $q_1$) -> synchRecv(S.R); currPort = S.R; currentLocation =  $q_2$;
    ::(currentLocation == $q_2$) -> synchRecv(S.cpr$_1$); currPort = S.cpr$_1$;  $q_3$;
    ::(currentLocation == $q_3$) -> send(B$_1$.R); send(B$_2$.R); recvAck(B$_1$.R); recvAck(B$_2$.R); currPort = S.S; currentLocation = $q_4$;
    ::(currentLocation == $q_4$) -> send(B$_1$.cpr$_1$); recvAck(B$_1$.cpr$_1$); currPort = S.cps$_1$ currentLocation = $q_5$;
    ::(currentLocation == $q_5$) -> 
      if
      ::recv(S.cpr$_2$) -> sendAck(S.cpr$_2$); currPort = S.cpr$_2$; currentLocation = $q_6$;
      ::recv(S.cpr$_3$) -> sendAck(S.cpr$_3$); currPort = S.cpr$_3$; currentLocation = $q_9$;
      fi;
    ::(currentLocation == $q_6$) -> synchRecv(S.R); currPort = S.R; currentLocation = $q_7$;
    ::(currentLocation == $q_7$) -> synchRecv(S.cpr$_4$); currPort = S.cpr$_4$; currentLocation = $q_8$;
    ::(currentLocation == $q_8$) -> send(B$_1$.R); send(B$_2$.R); recvAck(B$_1$.R); recvAck(B$_2$.R); currPort = S.S; currentLocation = $q_9$;
    ::(currentLocation == $q_9$) -> send(B$_2$.cpr$_2$); recvAck(B$_2$.cpr$_2$); currPort = S.cps$_2$; currentLocation = $q_{10}$;
    ::(currentLocation == $q_{10}$) ->
      if
      ::recv(S.cpr$_5$) -> sendAck(S.cpr$_5$); currPort = S.cpr$_5$; currentLocation = $q_{11}$
      ::recv(S.cpr$_6$) -> sendAck(S.cpr$_5$); currPort = S.cpr$_6$; currentLocation = $q_{14}$
      fi;   
    ::(currentLocation == $q_{11}$) -> synchRecv(S.R); currPort = S.R; currentLocation = $q_{12}$;
    ::(currentLocation == $q_{12}$) -> synchRecv(S.R); currPort = S.R; currentLocation = $q_{13}$;   
    ::(currentLocation == $q_{13}$) -> synchRecv(S.cpr$_7$); currPort = S.cpr$_7$; currentLocation = $q_{14}$;
    ::(currentLocation == $q_{14}$) -> currPort = S.E; currentLocation = end;
    ::(currentLocation == end) -> break;
    fi;
    updateCurrentState();
  od;
}
\end{lstlisting}
\texttt{updateCurrentState} is a macro that updates the current location and execute the location function of the current outgoing transition. The result of this computation would then be stored in the variable \texttt{value}.
%
\subsection{Verifying the Implementation}
%
We verify the generated implementation of the buying system against LTL~\cite{Pnueli77}\footnote{We recall the intuitive meaning of LTL operators: $\mathbf{G} \varphi$ (resp. $\mathbf{F} \varphi$, $\mathbf{X} \varphi$) stands for globally (resp. eventually, next) $\varphi$, and $\varphi_1 \mathbf{U} \varphi_2$ stands for $\varphi_1$ until $\varphi_2$.}  properties specifying its expected behavior.
In the following descriptions of properties, we prefix variables local to processes with the the name of the process.
\paragraph{Correct termination}
The correct termination property require that ``all processes terminate if any of them terminate".
Let the ports suffixed by \texttt{E} represent the termination interface/port of the corresponding process.
Moreover, we consider the following atomic propositions \texttt{currPort}$_1$ = \texttt{Buyer1.currPort}, \texttt{currPort$_2$} = \texttt{Buyer2.currPort}, \texttt{currPort$_3$} = \texttt{Bank.currPort}, and \texttt{currPort$_4$ = Seller.currPort}.
Then, correct termination can be expressed as the following LTL formula:
\[
\mathbf{G} \left(\bigvee_{i = 1}^{4} (\mathtt{currPort}_i = E_i) \implies \mathbf{F}\bigwedge_{i = 1}^{4} (\mathtt{currPort}_i = E_i) \right)
\] 
where $\mathtt{E_i}$ represents the ending interface of the appropriate process.
\paragraph{Absence of livelock}
Progress must be made towards termination (i.e., there are no cyclic paths with no work accomplished).
Intuitively, the system is in livelock state if the port \texttt{Bk.InfR} is used infinitely often along an execution path.
Therefore, specifying that the system is free of livelock can be modeled as the LTL formula:
\[
\mathbf{\neg} \big( \mathbf{GF}\left( \mathtt{Bank.currPort = Bk.InfR} \right) \big)
\]
\paragraph{Uniqueness of interface calls}
An interface should \textit{only be called once}.
In each run, money is only withdrawn once by each process.
Let the port \texttt{Bk.MS$_1$} (resp. \texttt{Bk.MS$_2$}) represent the withdrawal of money by process 1 (resp. process 2).
Then, specifying that money is withdrawn once per process can be expressed as the LTL formula:
\[
\bigwedge_{i=1}^2 \mathbf{G}((\mathtt{Bank.currPort = Bk.MS}_i) \implies \mathbf{XG} (\mathbf{\neg} \mathtt{Bank.currPort = Bk.MS}_i) )
\]
\paragraph{Correct transaction}
Money is only withdrawn \textit{after} either Buyer1 or Buyer 2 makes a request.
Let the ports \texttt{Bk.MS$_i$} be as above and let \texttt{B$_i$.MS} represent money transfer requests by Buyer $i$. Then specifying the order of execution is represented by the following LTL formula:
\[
\bigwedge_{i=1}^2 \mathbf{G} \big((\neg (\mathtt{Bank.currPort = Bk.MS}_i))\; \textbf{U}\; (\mathtt{B_i.currPort = B_i.MS}) \big)
\]
\section{Related Work}
\label{sec:rw}
%
%
Many coordination models exist to simplify the modeling of interactions in concurrent and distributed systems, such as in~\cite{AghaK99,bip2}.
Using these models requires the definition of the local behaviors of the processes and use of the communication model to implement the interactions between them.
This is in contrast to our case where we automatically synthesize the local code of the processes.

Moreover, in order to reason about the correctness of coordinated processes, Session types~\cite{BejleriY09,HondaYC08,BonelliC07,VallecilloVR06,GayVRGC10,CharalambidesDA16} and choreographies~\cite{TuostoG18} have been proposed to statically verify the implementations of communication protocols based on the following methodology: (1) define communication protocol between processes using a \emph{global protocol}; (2) automatically synthesize \emph{local types} which are the projection of global protocol w.r.t. processes; (3) develop the code of processes; (4) statically type-check the code of the processes w.r.t. local types.
Consequently, the distributed software follows the stipulated global protocol.
In our case, we automatically generate a more refined version of processes that embeds all the communication and synchronization logic as well as control-flows,  and which is correct-by-construction with respect to the global choreography. 
In~\cite{CarboneM13}, the authors present a deadlock-freedom by design method for choreographies communicating using multiparty asynchronous interactions. The method allows to efficiently verify and reason at the choreography level. Although, (1) the method is not concerned about synthesizing distributed implementation; and (2) the communication model only supports asynchronous interactions; using this approach can help us to reason and verify about our choreographies. Moreover, we can use a similar approach introduced in~\cite{ScalasY19} to efficiently verify our  choreographies.
In~\cite{LangeT12,LangeTY15}, the authors present a method to synthesize a global choreography from a set of local types.
The global view allows for the reasoning and analysis of distributed systems.
In our approach, we consider the inverse of that transformation, i.e., we create a template with all the necessary communication and control flows of the end-point processes starting from a global choreography. 

In~\cite{AttieD97,FrancezF91}, the authors introduce syntactic transformations to refine distributed system programs starting from high-level specifications.
In~\cite{AttieD97}, the proposed specification differs from our choreography model as it is not possible to express multiparty interactions, or guarded loop, which makes it impractical in the context of distributed systems.
In~\cite{FrancezF91}, the paper mainly targets multiparty interactions, where the main objective is to loosening synchronous multiparty interaction while preserving its semantics.
In our case, as we automatically synthesize code for multiply interactions, there is no need for  loosening technique. Add to that, we also support asynchronous ports that allow to loosening interactions.
Additionally, in~\cite{AttieD97,FrancezF91}, it is not clear how to automatically generate code from the refined programs.

BPMN~\cite{bpmn} (Business Process Model and Notation) is an industry standard that allows to model process choreographies.
An extension of BPMN was introduced in~\cite{HofreiterH08,PBFW15} to automatically derive a local choreography from a global one.
Nonetheless, the extension only considers exchange of messages and does not formally define other composition operators such as synchronous multiparty communications, parallelism, choice,  sequential and loop.
The method proposed in~\cite{NikajWM19} allows to derive RESTful choreographies from process choreographies, whereas in this paper we synthesize the code of the processes given global choreography. Moreover, the model is restricted to RESTful architecture.
In~\cite{GudemannPSY16}, the authors introduce a framework for the verification and design of choreographies, however, the communication model only allows for one send and one receive per interaction. 








\section{Conclusion and Future Work}
\label{sec:conclusion}
%
%
\paragraph{Conclusion}
This paper deals with the synthesis of distributed implementations of local processes (control flows, synchronization, notification, acknowledgment, computations embedding), starting from a global choreography.
The method presented in this paper allows one to automatically verify the communication protocols and drastically simplify the synthesis of the distributed implementation. 
Moreover, the language is used to model a real case study provided by Murex S.A.L. services industry. 
We used the choreography language and the method to synthesize actual micro-services architectures.
The synthesized micro-services can be verified against any Linear Temporal Logic formula thanks to a translation to Promela.
We illustrated the translation and the verification on a simplified version of an application at Murex for which we synthesized the micro-service implementation.
\paragraph{Future work}
Future work comprises several directions.
First, we consider augmenting our choreography model by adding fault-tolerance primitives.
That is, we aim to specify the number of replicas of each process and automatically embed a consensus protocol between them such as Paxos~\cite{paxos} or Raft~\cite{raft}. 
Second, we consider integrating our framework with Spring Boot to allow for the automatic generation of RESTful web services starting from global choreography.
Third, we consider augmenting our code generation with features provided by \emph{Istio}~\cite{istio} and \emph{Linkerd}~\cite{linkerd}, which are used for routing, failure handling, service discovery, the integration of micro-services, the traffic-flow management and enforcing policies. 
Fourth, we consider defining a specific model-checker for our distributed component-based framework.  
Finally, we consider using complementary verification techniques operating at runtime such as runtime verification~\cite{lncs/10457} and runtime enforcement \cite{Falcone10} for which we defined approaches in the case of non-distributed component-based systems~\cite{FalconeJNBB15,FalconeJ17}.
\subsection*{Acknowledgment}
The work presented in this paper is supported by the Murex Grant Award 103456 and University Research Board at the American University of Beirut.
%
%
\bibliographystyle{elsarticle-harv}
\bibliography{biblio}

\begin{thebibliography}{33}
\expandafter\ifx\csname natexlab\endcsname\relax\def\natexlab#1{#1}\fi
\expandafter\ifx\csname url\endcsname\relax
  \def\url#1{\texttt{#1}}\fi
\expandafter\ifx\csname urlprefix\endcsname\relax\def\urlprefix{URL }\fi

\bibitem[{Agha and Kim(1999)}]{AghaK99}
Agha, G.~A., Kim, W., 1999. Actors: {A} unifying model for parallel and
  distributed computing. Journal of Systems Architecture 45~(15), 1263--1277.
\newline\urlprefix\url{https://doi.org/10.1016/S1383-7621(98)00067-8}

\bibitem[{Attie and Das(1997)}]{AttieD97}
Attie, P.~C., Das, C., 1997. Automating the refinement of specifications for
  distributed systems via syntactic transformations. Int. J. Systems Science
  28~(11), 1129--1144.

\bibitem[{Bartocci and Falcone(2018)}]{lncs/10457}
Bartocci, E., Falcone, Y. (Eds.), 2018. Lectures on Runtime Verification -
  Introductory and Advanced Topics. Vol. 10457 of Lecture Notes in Computer
  Science. Springer.
\newline\urlprefix\url{https://doi.org/10.1007/978-3-319-75632-5}

\bibitem[{Basu et~al.(2011)Basu, Bensalem, Bozga, Combaz, Jaber, Nguyen, and
  Sifakis}]{bip2}
Basu, A., Bensalem, S., Bozga, M., Combaz, J., Jaber, M., Nguyen, T., Sifakis,
  J., 2011. Rigorous component-based system design using the {BIP} framework.
  {IEEE} Software 28~(3), 41--48.

\bibitem[{Bejleri and Yoshida(2009)}]{BejleriY09}
Bejleri, A., Yoshida, N., 2009. Synchronous multiparty session types. Electr.
  Notes Theor. Comput. Sci. 241, 3--33.

\bibitem[{Bonakdarpour et~al.(2012)Bonakdarpour, Bozga, Jaber, Quilbeuf, and
  Sifakis}]{BonakdarpourBJQS12}
Bonakdarpour, B., Bozga, M., Jaber, M., Quilbeuf, J., Sifakis, J., 2012. A
  framework for automated distributed implementation of component-based models.
  Distributed Computing 25~(5), 383--409.

\bibitem[{Bonelli and Compagnoni(2007)}]{BonelliC07}
Bonelli, E., Compagnoni, A.~B., 2007. Multipoint session types for a
  distributed calculus. In: Trustworthy Global Computing, Third Symposium,
  {TGC} 2007, Sophia-Antipolis, France, November 5-6, 2007, Revised Selected
  Papers. pp. 240--256.

\bibitem[{Carbone and Montesi(2013)}]{CarboneM13}
Carbone, M., Montesi, F., 2013. Deadlock-freedom-by-design: multiparty
  asynchronous global programming. In: The 40th Annual {ACM} {SIGPLAN-SIGACT}
  Symposium on Principles of Programming Languages, {POPL} '13, Rome, Italy -
  January 23 - 25, 2013. pp. 263--274.
\newline\urlprefix\url{https://doi.org/10.1145/2429069.2429101}

\bibitem[{Charalambides et~al.(2016)Charalambides, Dinges, and
  Agha}]{CharalambidesDA16}
Charalambides, M., Dinges, P., Agha, G.~A., 2016. Parameterized, concurrent
  session types for asynchronous multi-actor interactions. Sci. Comput.
  Program. 115-116, 100--126.

\bibitem[{Falcone(2010)}]{Falcone10}
Falcone, Y., 2010. You should better enforce than verify. In: Barringer, H.,
  Falcone, Y., Finkbeiner, B., Havelund, K., Lee, I., Pace, G.~J., Rosu, G.,
  Sokolsky, O., Tillmann, N. (Eds.), Runtime Verification - First International
  Conference, {RV} 2010, St. Julians, Malta, November 1-4, 2010. Proceedings.
  Vol. 6418 of Lecture Notes in Computer Science. Springer, pp. 89--105.
\newline\urlprefix\url{https://doi.org/10.1007/978-3-642-16612-9\_9}

\bibitem[{Falcone and Jaber(2017)}]{FalconeJ17}
Falcone, Y., Jaber, M., 2017. Fully automated runtime enforcement of
  component-based systems with formal and sound recovery. {STTT} 19~(3),
  341--365.
\newline\urlprefix\url{https://doi.org/10.1007/s10009-016-0413-6}

\bibitem[{Falcone et~al.(2015)Falcone, Jaber, Nguyen, Bozga, and
  Bensalem}]{FalconeJNBB15}
Falcone, Y., Jaber, M., Nguyen, T., Bozga, M., Bensalem, S., 2015. Runtime
  verification of component-based systems in the {BIP} framework with
  formally-proved sound and complete instrumentation. Software and System
  Modeling 14~(1), 173--199.
\newline\urlprefix\url{https://doi.org/10.1007/s10270-013-0323-y}

\bibitem[{Francez and Forman(1991)}]{FrancezF91}
Francez, N., Forman, I.~R., 1991. Synchrony loosening transformations for
  interacting processes. In: {CONCUR} '91, 2nd International Conference on
  Concurrency Theory, Amsterdam, The Netherlands, August 26-29, 1991,
  Proceedings. pp. 203--219.

\bibitem[{Francez and Forman(2018)}]{hpcs4padjaber}
Francez, N., Forman, I.~R., 2018. From global choreography to efficient
  distributed implementation. In: {HPCS - 4PAD} International Symposium on
  Formal Approaches to Parallel and Distributed Systems.

\bibitem[{Gay et~al.(2010)Gay, Vasconcelos, Ravara, Gesbert, and
  Caldeira}]{GayVRGC10}
Gay, S.~J., Vasconcelos, V.~T., Ravara, A., Gesbert, N., Caldeira, A.~Z., 2010.
  Modular session types for distributed object-oriented programming. In:
  Proceedings of the 37th {ACM} {SIGPLAN-SIGACT} Symposium on Principles of
  Programming Languages, {POPL} 2010, Madrid, Spain, January 17-23, 2010. pp.
  299--312.

\bibitem[{G{\"{u}}demann et~al.(2016)G{\"{u}}demann, Poizat, Sala{\"{u}}n, and
  Ye}]{GudemannPSY16}
G{\"{u}}demann, M., Poizat, P., Sala{\"{u}}n, G., Ye, L., 2016. Verchor: {A}
  framework for the design and verification of choreographies. {IEEE} Trans.
  Services Computing 9~(4), 647--660.
\newline\urlprefix\url{https://doi.org/10.1109/TSC.2015.2413401}

\bibitem[{Hofreiter and Huemer(2008)}]{HofreiterH08}
Hofreiter, B., Huemer, C., 2008. A model-driven top-down approach to
  inter-organizational systems: From global choreography models to executable
  {BPEL}. In: 10th {IEEE} International Conference on E-Commerce Technology
  {(CEC} 2008) / 5th {IEEE} International Conference on Enterprise Computing,
  E-Commerce and E-Services {(EEE} 2008), July 21-14, 2008, Washington, DC,
  {USA}. pp. 136--145.
\newline\urlprefix\url{https://doi.org/10.1109/CECandEEE.2008.129}

\bibitem[{Holzmann(1997)}]{promelaspin}
Holzmann, G.~J., 1997. The model checker {SPIN}. {IEEE} Trans. Software Eng.
  23~(5), 279--295.
\newline\urlprefix\url{https://doi.org/10.1109/32.588521}

\bibitem[{Honda et~al.(2008)Honda, Yoshida, and Carbone}]{HondaYC08}
Honda, K., Yoshida, N., Carbone, M., 2008. Multiparty asynchronous session
  types. In: Proceedings of the 35th {ACM} {SIGPLAN-SIGACT} Symposium on
  Principles of Programming Languages, {POPL} 2008, San Francisco, California,
  USA, January 7-12, 2008. pp. 273--284.

\bibitem[{Istio(.)}]{istio}
Istio, . \url{https://github.com/istio/istio/}.

\bibitem[{Lamport(2001)}]{paxos}
Lamport, L., December 2001. Paxos made simple, 51--58.
\newline\urlprefix\url{https://www.microsoft.com/en-us/research/publication/paxos-made-simple/}

\bibitem[{Lange and Tuosto(2012)}]{LangeT12}
Lange, J., Tuosto, E., 2012. Synthesising choreographies from local session
  types. In: {CONCUR} 2012 - Concurrency Theory - 23rd International
  Conference, {CONCUR} 2012, Newcastle upon Tyne, UK, September 4-7, 2012.
  Proceedings. pp. 225--239.

\bibitem[{Lange et~al.(2015)Lange, Tuosto, and Yoshida}]{LangeTY15}
Lange, J., Tuosto, E., Yoshida, N., 2015. From communicating machines to
  graphical choreographies. In: Proceedings of the 42nd Annual {ACM}
  {SIGPLAN-SIGACT} Symposium on Principles of Programming Languages, {POPL}
  2015, Mumbai, India, January 15-17, 2015. pp. 221--232.

\bibitem[{Linkerd(.)}]{linkerd}
Linkerd, . \url{https://linkerd.io}.

\bibitem[{Meyer et~al.(2015)Meyer, Pufahl, Batoulis, Fahland, and
  Weske}]{PBFW15}
Meyer, A., Pufahl, L., Batoulis, K., Fahland, D., Weske, M., 2015. Automating
  data exchange in process choreographies. Inf. Syst. 53, 296--329.
\newline\urlprefix\url{https://doi.org/10.1016/j.is.2015.03.008}

\bibitem[{Murex(.)}]{murex}
Murex, . \url{https://www.murex.com}.

\bibitem[{Nikaj et~al.(2019)Nikaj, Weske, and Mendling}]{NikajWM19}
Nikaj, A., Weske, M., Mendling, J., 2019. Semi-automatic derivation of restful
  choreographies from business process choreographies. Software and System
  Modeling 18~(2), 1195--1208.
\newline\urlprefix\url{https://doi.org/10.1007/s10270-017-0653-2}

\bibitem[{OMG and Notation~(BPMN)(2011)}]{bpmn}
OMG, B. P.~M., Notation~(BPMN), V.~., 2011.
  \url{http://www.omg.org/spec/BPMN/2.0/}.

\bibitem[{Ongaro and Ousterhout(2014)}]{raft}
Ongaro, D., Ousterhout, J.~K., 2014. In search of an understandable consensus
  algorithm. In: 2014 {USENIX} Annual Technical Conference, {USENIX} {ATC} '14,
  Philadelphia, PA, USA, June 19-20, 2014. pp. 305--319.

\bibitem[{Pnueli(1977)}]{Pnueli77}
Pnueli, A., 1977. The temporal logic of programs. In: 18th Annual Symposium on
  Foundations of Computer Science, Providence, Rhode Island, USA, 31 October -
  1 November 1977. {IEEE} Computer Society, pp. 46--57.

\bibitem[{Scalas and Yoshida(2019)}]{ScalasY19}
Scalas, A., Yoshida, N., 2019. Less is more: multiparty session types
  revisited. {PACMPL} 3~({POPL}), 30:1--30:29.
\newline\urlprefix\url{https://doi.org/10.1145/3290343}

\bibitem[{Tuosto and Guanciale(2018)}]{TuostoG18}
Tuosto, E., Guanciale, R., 2018. Semantics of global view of choreographies. J.
  Log. Algebr. Meth. Program. 95, 17--40.
\newline\urlprefix\url{https://doi.org/10.1016/j.jlamp.2017.11.002}

\bibitem[{Vallecillo et~al.(2006)Vallecillo, Vasconcelos, and
  Ravara}]{VallecilloVR06}
Vallecillo, A., Vasconcelos, V.~T., Ravara, A., 2006. Typing the behavior of
  software components using session types. Fundam. Inform. 73~(4), 583--598.

\end{thebibliography}
%
%
\end{document}